\documentclass[graybox]{svmult}

\usepackage{mathptmx}       
\usepackage{helvet}         
\usepackage{courier}        
\usepackage{type1cm}        
\usepackage{makeidx}         
\usepackage{graphicx}        
\usepackage{multicol}        
\usepackage[bottom]{footmisc}
\usepackage{hyperref}        
\usepackage{soul}            
\hypersetup{colorlinks=true,urlcolor=blue}
\usepackage{natbib}
\usepackage{aas_macros}
\usepackage{amsmath}
\usepackage{amssymb}	
\makeindex             

\usepackage{xcolor}
\usepackage[T1]{fontenc}
\usepackage{booktabs}
\usepackage{array}

\newcommand{\Msol}{\ensuremath{\text{M}_{\odot}}}
\newcommand{\keV}{\ensuremath{\text{~keV}}}

\newcommand{\Rf}{\ensuremath{R_{\text{500}}}}
\newcommand{\Mf}{\ensuremath{M_{\text{500}}}}
\newcommand{\Mgas}{\ensuremath{M_{\text{gas}}}}

\newcommand{\MHE}{\ensuremath{M_{\text{HE}}}}
\newcommand{\fgas}{\ensuremath{f_{\text{gas}}}}
\newcommand{\rhog}{\ensuremath{\rho_{\text{gas}}}}
\newcommand{\Tx}{\ensuremath{T_{\text{X}}}}
\newcommand{\Lx}{\ensuremath{L_{\text{X}}}}
\newcommand{\Lbol}{\ensuremath{L_{\text{X,bol}}}}
\newcommand{\Lsoft}{\ensuremath{L_{\text{X,soft}}}}
\newcommand{\Lce}{\ensuremath{L_{\text{X,ce}}}}
\newcommand{\Yx}{\ensuremath{Y_{\text{X}}}}
\newcommand{\Mstar}{\ensuremath{M_\star}}
\newcommand{\nel}{\ensuremath{n_{\text{e}}}}
\newcommand{\np}{\ensuremath{n_{\text{p}}}}

\def\MgM{M$_{\rm gas}$--M}
\def\TM{T$_{\rm X}$--M}
\def\MT{M--T$_{\rm X}$}
\def\LM{L$_{\rm X}$--M}
\def\YM{Y$_{\rm X}$--M}
\def\MY{M--Y$_{\rm X}$}
\def\LT{L$_{\rm X}$--T$_{\rm X}$}
\def\LbolT{L$_{\rm X,bol}$--T$_{\rm X}$}

\newcommand{\Amgm}{\ensuremath{A_{\text{mM}}}}
\newcommand{\Bmgm}{\ensuremath{B_{\text{mM}}}}
\newcommand{\gmgm}{\ensuremath{\gamma_{\text{mM}}}}
\newcommand{\Atm}{\ensuremath{A_{\text{TM}}}}
\newcommand{\Btm}{\ensuremath{B_{\text{TM}}}}
\newcommand{\gtm}{\ensuremath{\gamma_{\text{TM}}}}
\newcommand{\Alm}{\ensuremath{A_{\text{LM}}}}
\newcommand{\Blm}{\ensuremath{B_{\text{LM}}}}
\newcommand{\glm}{\ensuremath{\gamma_{\text{LM}}}}
\newcommand{\Aym}{\ensuremath{A_{\text{YM}}}}
\newcommand{\Bym}{\ensuremath{B_{\text{YM}}}}
\newcommand{\gym}{\ensuremath{\gamma_{\text{YM}}}}
\newcommand{\Alt}{\ensuremath{A_{\text{LT}}}}
\newcommand{\Blt}{\ensuremath{B_{\text{LT}}}}
\newcommand{\glt}{\ensuremath{\gamma_{\text{LT}}}}
\newcommand{\Akt}{\ensuremath{A_{\text{KT}}}}
\newcommand{\Bkt}{\ensuremath{B_{\text{KT}}}}
\newcommand{\gkt}{\ensuremath{\gamma_{\text{KT}}}}

\newcommand{\Chandra}{\textit{Chandra}}

\newcommand{\XMM}{XMM-\textit{Newton}}

\begin{document}
\title*{Scaling relations of clusters and groups, and their evolution}
\author{Lorenzo Lovisari \thanks{Corresponding author; Both authors contributed equally to this work.} and Ben J. Maughan}
\institute{Lorenzo Lovisari \at INAF - Osservatorio di Astrofisica e Scienza dello Spazio di Bologna, via Piero Gobetti 93/3, 40129 Bologna, Italia; \\
Center for Astrophysics | Harvard \& Smithsonian, 60 Garden Street, Cambridge, MA 02138, USA; \email{lorenzo.lovisari@inaf.it}
\and Ben J. Maughan \at H. H. Wills Physics Laboratory, University of Bristol, Tyndall Ave, Bristol BS8 1TL, UK;\\ \email{ben.maughan@bristol.ac.uk }
\vspace{2pt}
}
%
%
\maketitle
%

\abstract{X-ray observations of the hot intra-cluster medium (ICM) in galaxy groups and clusters provide quantities such as their gas mass, X-ray luminosity, and temperature.
The analysis of the scaling relations between these observable properties gives considerable insight into the physical processes taking place in the ICM.
Furthermore, an understanding of the scaling relations between ICM properties and the total cluster mass is essential for cosmological studies with clusters.
For these reasons, the X-ray scaling relations of groups and clusters have been a major focus of research over the past several decades.
In this Chapter, after presenting the expectations from the self–similar model, based on the assumption that only gravity drives the evolution of the ICM, we discuss how the processes of gas cooling and non–gravitational heating are believed to be responsible for the observed deviations from the self-similar scenario.
We also describe important complications that must be considered when measuring and interpreting the scaling relations.}

\section{Keywords}
galaxies: clusters: general – galaxies: clusters: intracluster medium – cosmology: observations – X-rays: galaxies: clusters

\bigskip

\section{Introduction}
Clusters of galaxies are among the most significant cosmic structures, both in terms of their mass and scale, but also for the depth of insight they can yield into both cosmology and astrophysics. 
As their name implies, they appear in optical wavelengths as collections of galaxies numbering anywhere between a few to a few thousand. 
These galaxies, however, account for just a few percent of the mass of a galaxy cluster. 
The dominant baryonic component of a cluster is a hot, tenuous and almost fully ionised gas which pervades the entire cluster and is known as the intra-cluster medium (ICM). 
The ICM accounts for $10$--$20\%$ of the mass of a cluster, and is the source of the bright diffuse X-ray emission that is characteristic of galaxy clusters. 
The gravitational potential is dominated by the dark matter which constitutes $80$--$90\%$ of the total mass of a cluster.

Galaxy clusters span a mass range of approximately $10^{13}$--10$^{15}\Msol$, although systems at the low mass end of the range are often referred to as galaxy groups. 
There is no precise definition of the boundary between groups and clusters, but a mass of $10^{14}\Msol$ is a reasonable working definition. 
The lower mass limit of galaxy groups is not sharply defined, since an isolated massive galaxy with a significant amount of dark matter can be regarded in some ways as a galaxy group with just one member.

In the current bottom-up scenario for the formation of cosmic structure, tiny density fluctuations are amplified by gravity  to create the massive, dark matter dominated structures we observe today. 
Thus, to a first approximation, clusters of galaxies are simple objects whose properties are defined only by their mass and redshift. 
This leads to simple power-law correlations between different observable properties of clusters, known as scaling relations. 
The study of these scaling relations has great utility in advancing our understanding of both astrophysics and cosmology. 
The correlations between observable properties of clusters and their masses provide the possibility of using easily measured properties as proxies for cluster masses, which are the essential quantity when using galaxy clusters to derive cosmological constraints. 
Meanwhile, the deviations of real clusters from the simple theoretical scaling relations are diagnostics of important astrophysical processes which shape the mutual growth and evolution of the different components of clusters. 
These two applications are not independent. 
For example, the impact of astrophysical processes on the scatter and evolution of scaling relations must be understood if they are to be used to provide masses for cosmological studies; neglecting this could mask or mimic an interesting cosmological signal in the data.

Almost as soon as clusters of galaxies were detected in X-rays, the correlations between their X-ray properties were investigated, revealing clear scaling relations between ICM properties (\citealt{mit79})\footnote{See \cite{sar86} for a comprehensive review of the X-ray properties of galaxy clusters.}. 
This early work set the scene for decades of more detailed investigations to explore the astrophysics inside the scaling relations, powered by a succession of ever-improving X-ray telescopes.

The overall picture that has emerged is that the scaling of cluster properties is dominated by mass, but for lower mass systems other effects become increasingly important, causing departures from simple scaling relations. 
These include the radiative cooling of the ICM; energy injected into the ICM from supernovae and active galactic nuclei (AGN) in the member galaxies; and stochastic and transient perturbations of cluster properties due to mergers with other groups and clusters. However, our understanding of these processes is far from complete.

Alongside the improvements in observing facilities, developments in numerical simulations of clusters in cosmological volumes have been critical for understanding the scaling relations. 
With increasing computing power, the resolution and sophistication of simulations has progressed from ``N-body'' simulations which model only the gravitational interactions of dark matter particles, to full hydrodynamical modelling of dark matter and baryons in cosmological volumes. 
However, including astrophysical processes in cosmological simulations is non-trivial because of the significantly different scales at work: the accreting AGN ($<$1 pc), its host galaxy ($\sim$ several kpc), the galaxy cluster ($\sim$1 Mpc), and the region that will collapse to form the final system ($\sim$10 Mpc). Important improvements have been achieved thanks to advancement in the treatment of physical processes acting below the resolution scale of the simulation (usually referred to as ``sub-grid'' models).

In this Chapter we will focus on the X-ray scaling relations of groups and clusters of galaxies. We will first present the theoretical background of the scaling of ICM properties with mass, and the astrophysical processes that are believed to be responsible for shaping the scaling relations. We will also describe a number of important biases and other issues that should be considered when measuring the scaling relations. We will review the current observational constraints on the scaling relations.  The implications of these observational results for our understanding of the astrophysical processes responsible for shaping the scaling relations will be discussed, including comparisons with the results from numerical simulations. Finally, we will conclude with an outlook on how new X-ray telescopes will help push forward the frontiers of this field.

\section{Theoretical background}

\subsection{The X-ray emission from clusters of galaxies}
\label{sec:x-ray-emission}
The fundamental physical properties of the ICM are its temperature ($\Tx$), density (usually expressed in terms of the number density of free electrons, $\nel$, and protons, $\np$), and abundance of heavy elements ($Z$, defined more precisely below). The main cooling process for the ICM is the emission of X-ray radiation, whose emissivity (i.e., the energy emitted per unit time and volume at some photon energy) is equal to
\begin{equation}
  \label{eq:emis}
    \epsilon=\nel \np \Lambda(\Tx,Z)
\end{equation}
where $\Lambda$ is known as the cooling function. The form of $\Lambda(\Tx,Z)$ depends on the details of the emission mechanisms, which are calculated by complex emission codes. To a good approximation, $\Lambda(\Tx,Z)$ is given by the sum of a thermal bremsstrahlung continuum and emission lines from collisionally excited metals in the ICM. The metals in the ICM contribute to the emission both through the obvious lines and also the continuum component (e.g., \citealt{2008SSRv..134..155K},  \citealt{boh10}).

The temperature of the ICM can be measured by fitting models to the emitted X-ray spectrum. Conventionally, X-ray spectral models parametrise the temperature as $kT$ where $k$ is the Boltzmann constant. X-ray temperatures are thus usually reported in units of keV, with values ranging from about $1\keV$ for low mass groups up to about $15\keV$ for very massive clusters.

The ICM is composed predominantly of fully ionised hydrogen and helium, with smaller amounts of heavier elements (whose ionisation states depend on the ICM properties). The heavy element (or metal) abundance in the ICM is usually measured relative to Solar abundances $Z_\odot$. It is common to assume that the relative abundances of elements heavier than helium is the same as that in the Sun. The abundance measurement is then the ratio of metals to hydrogen in the ICM compared to that ratio in the Sun. For example, a typical ICM abundance of $Z=0.3Z_\odot$ means that the ICM has $30\%$ as much metals relative to hydrogen compared to the Sun (see Mernier et al. - Chapter: {\it Chemical enrichment in groups and clusters} of this volume, for an extensive discussion on metal abundances in groups and clusters).

For typical ICM metal abundances, then densities of free electrons and protons are related by $\nel\approx1.2\np$. The mass density of the ICM, $\rhog$ can be derived from the number densities via the mean atomic mass per particle, $\mu\approx0.6$ for typical ICM composition.

\begin{figure}[t]
  \centering
  \includegraphics[width = 0.95\textwidth]{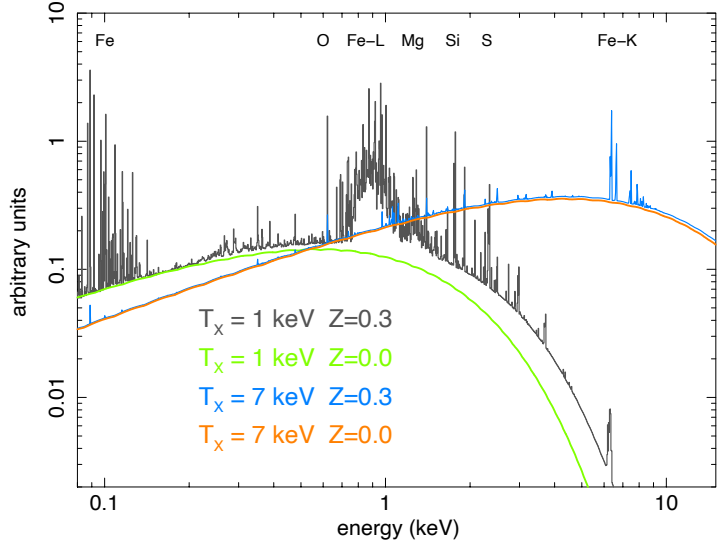}
  \caption{Model X-ray spectra for clusters of different temperatures,
    assuming $Z=0.3$. For comparison we show also the spectra with
    zero metallicity. The elements and ionisation states responsible
    for some of the most prominent emissions lines are indicated.}
  \label{fig:spectra}
\end{figure}

For ICM temperatures above about $3\keV$, the bremsstrahlung component dominates the X-ray spectrum, but at lower temperatures (i.e., lower mass clusters), other mechanisms (primarily line emission from collisionally excited ions) make increasingly important contributions to the X-ray emission (e.g., \citealt{lov21}). This is illustrated in Figure \ref{fig:spectra}, which shows the enhanced contribution of line emission in lower temperature clusters.

The X-ray luminosity of the ICM is given by integrating $\epsilon$ over the X-ray energy band being considered, and over the volume of the cluster\footnote{The volume is generally assumed to be spherical. 
However, the ICM is optically thin so the luminosity observed in a circular aperture of projected radius $R$ contains emission from ICM at larger radii in a cylinder projected along the line of sight. 
This additional contribution to the luminosity is  usually neglected because the emission is dominated by the denser ICM at smaller radii due to the $n_e$ and $n_p$ dependence.}. 
The X-ray emission from the ICM and the sensitivity of most X-ray observatories both peak around $\sim$1--2$\keV$, so it is often
convenient to measure X-ray luminosities in a ``soft'' energy band, $\Lsoft$ (0.1--2.4$\keV$ and 0.5--2$\keV$ are conventional)\footnote{We note that since clusters are generally observed at non-negligible redshifts, a cluster's X-ray flux must be ``k-corrected'' from the observer's frame to the desired energy band in the cluster's rest frame when calculating its luminosity (see, e.g., \citealt{1998ApJ...495..100J}).}. 
Alternatively, the bolometric X-ray luminosity, $\Lbol$, is given by integrating the emission across the whole X-ray band (the exact limits don't matter since the emissivity drops to zero at high and low energies, but the range 0.01--100 $\keV$ is commonly used).

In addition to the luminosity, X-ray observations allow one to easily derive the gas temperature and the emission measure\footnote{There is some ambiguity over this term in the literature. The term "emission integral" may also be used to refer to this quantity, or sometimes to refer to the related quantity of the integral of $n_e n_p$ along the line of sight.} (EM=$\int n_e n_p dV$) from spectral fits. The latter quantity is proportional to the normalisation of standard ICM spectral models and provides the means to measure the ICM density.

Given data of sufficient quality, it is possible to measure the temperature and density structure of the ICM. For example, radial profiles $\Tx(r)$ and $\rhog(r)$ can be recovered assuming spherical symmetry and accounting for projected emission along the line of sight. It is then possible to estimate the total gravitating mass of a cluster under the assumption that the ICM is in hydrostatic equilibrium (HE). The hydrostatic mass contained within a radius $r$ is given by
\begin{align}
    \MHE(<r) & = - \frac{k\Tx r}{G\mu m_p}\left[ \frac{\text{d} \log \rhog}{\text{d} \log r} + \frac{\text{d} \log \Tx}{\text{d} \log r}\right] .
\end{align}

\subsection{Self-similarity}
\label{sec:self-similarity}

Under some simplifying assumptions (that galaxy clusters grow from scale-free initial conditions, form recently in a single spherical collapse, and their energy content is only obtained from the gravitational potential in this collapse), \citet{kai86} showed that clusters should be self-similar objects (see \citealt{boh12}, for a more recent discussion of the self-similar model). 
This means that all clusters of the same mass and at the same redshift should be identical, and that their properties should scale simply with mass and redshift.

In this self-similar model, the collapsed region that formed the cluster is virialised (i.e., the kinetic energy of its components have obtained an equilibrium with the gravitational potential, obeying the virial theorem). 
One can then imagine a notional spherical surface separating the virialised material from the surroundings. 
The radius of this imaginary surface is known as the virial radius, and sets the spatial scale of clusters in the self-similar model.

The virial radius may be expressed in terms of an overdensity, $\Delta$, relative to the critical density of the Universe at the redshift of the cluster, $\rho_c(z)$, where
\begin{align}
  \rho_c(z) = \frac{3 H(z)^2}{8\pi G}.
\end{align}
Here, $H(z)$ is the Hubble parameter, which can be expressed in terms
of its present-day value as $H(z)=E(z)H_0$, where
\begin{align}
E(z) = \sqrt{\Omega_\text{M}(1+z)^3 + \Omega_k
  (1+z)^2 + \Omega_\Lambda},
\end{align}
with $\Omega_\text{M}$ and $\Omega_\Lambda$ being the density parameters associated to the non-relativistic matter and to the cosmological constant respectively, and $\Omega_k=(1-\Omega_\text{M}-\Omega_\Lambda)$ defines the curvature of the Universe and is equal to zero for a flat cosmology. 
Since in an expanding Universe the Hubble parameter $H$ is a function of time, then the critical density also changes with time (i.e., the value of $\rho_c$ increases with redshift). 

In general, we define an overdensity radius $R_\Delta$ such that
\begin{align}
\label{eq:Rdelta}
  \frac{M_\Delta}{\frac{4}{3}\pi R_\Delta^3} = \Delta \rho_c(z),
\end{align}
where $M_\Delta$ is the mass enclosed within a sphere of radius $R_\Delta$. 
In a flat, matter-dominated Universe (i.e., $\Omega_\text{M}$=1 and $\Omega_\Lambda$=0), the virial radius corresponds to $\Delta=18\pi^2\approx178$, but in general the value and redshift-dependence of the overdensity within the virial radius depend on the assumed cosmology (e.g., for $\Omega_\text{M}$=0.3 and $\Omega_\Lambda$=0.7 using Eq. 6 of \citealt{bry98} one finds $\Delta_{vir}(z=0)\approx100$). 
To a good approximation, however, clusters can be considered self-similar when they are scaled to a constant overdensity radius at all redshifts. 
In the case of X-ray observations of clusters, it is common to adopt $\Delta=500$, since $\Rf$ is the radius to which ICM properties can be measured in typical observations with the current
instruments\footnote{When galaxy cluster properties are used for cosmological studies (or any application in which the cosmological model is free to vary), one must take into account the various ways in which the cosmological model impacts the measurements. The most obvious are the conversion of observed quantities to physical quantities (i.e., angles to distances, and fluxes to luminosities). However, the use of an overdensity radius to measure cluster properties also introduces a cosmology dependence due to the cosmology-dependence of $\rho_c(z)$.}. An extensive description of the spherical collapse model and of its parameters can be found in Clerc et al. - Chapter: {\it X-ray cluster cosmology} of this volume.

\begin{figure}[t]
  \centering
  \includegraphics[width = 0.32\textwidth]{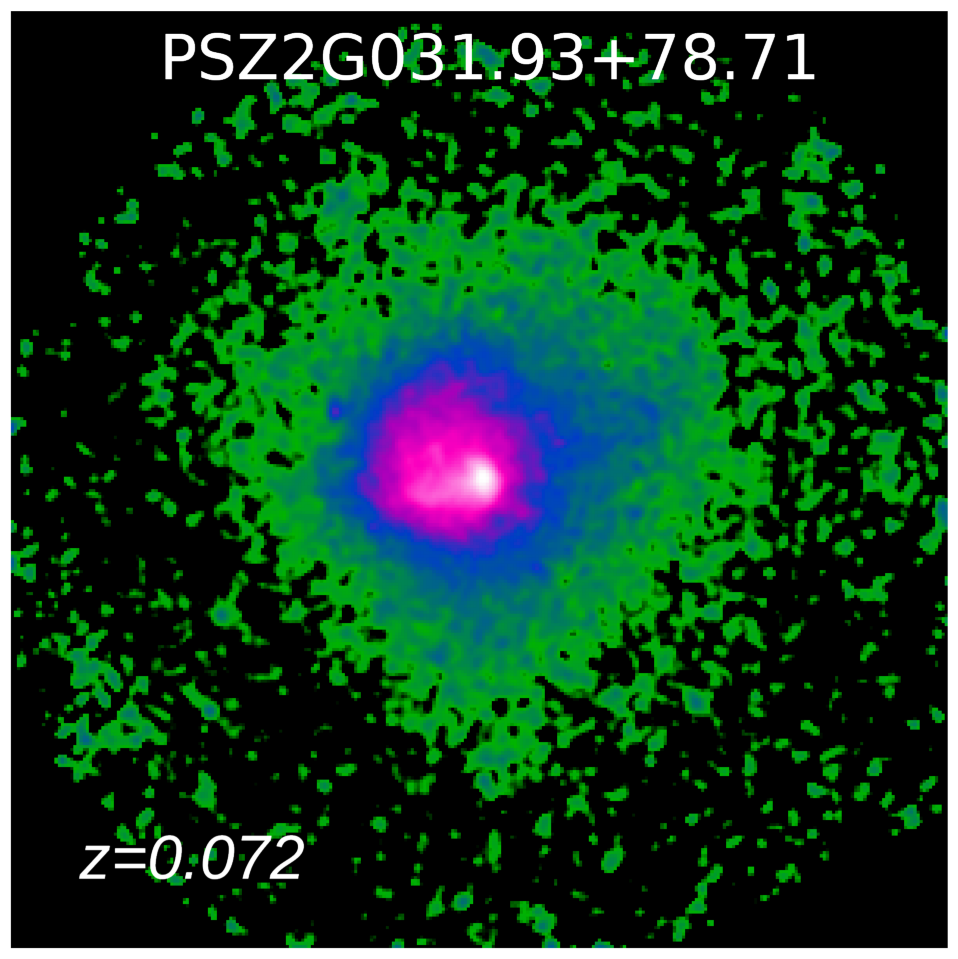}\,
  \includegraphics[width = 0.32\textwidth]{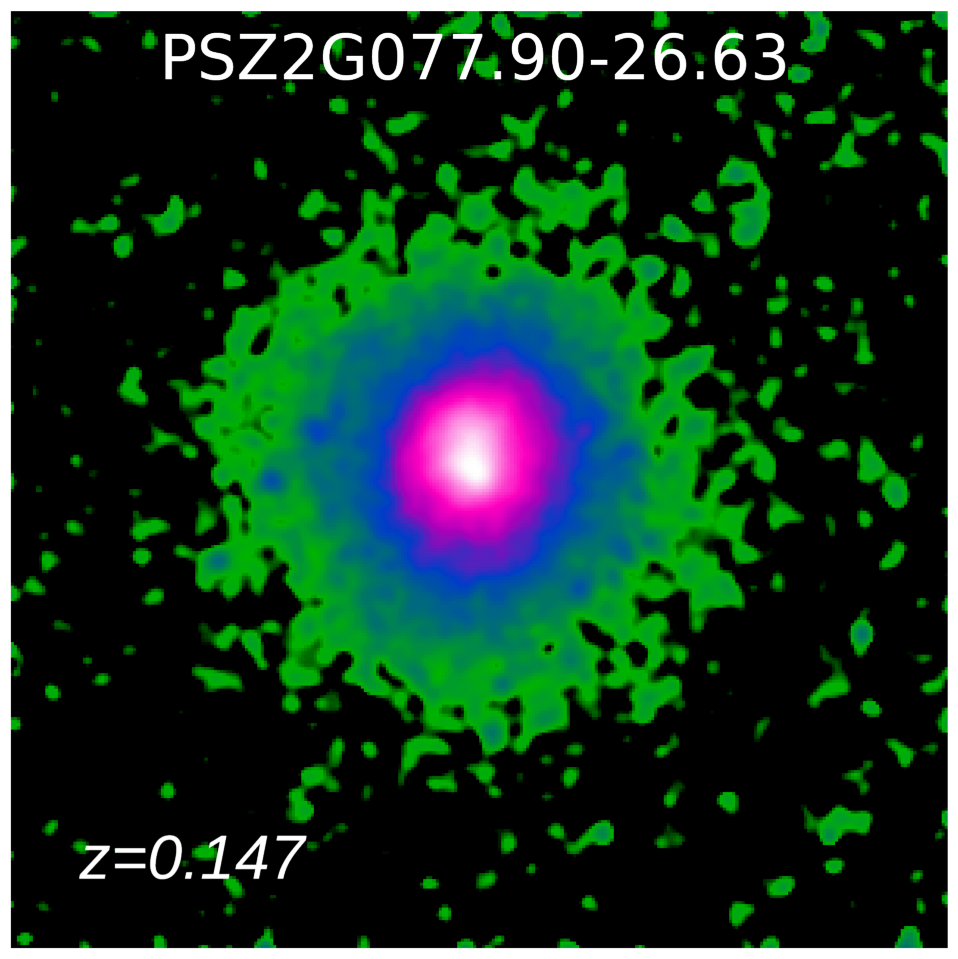}\,
  \includegraphics[width = 0.32\textwidth]{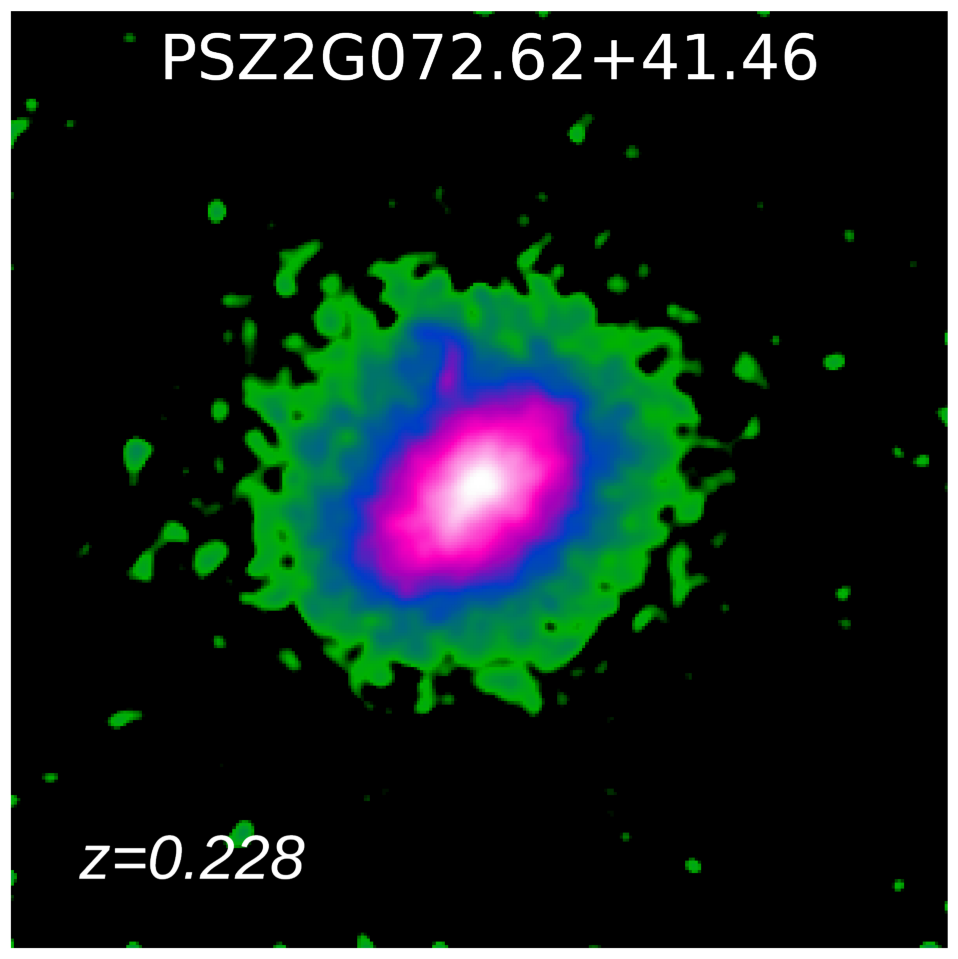}
  \caption{Smoothed \XMM\ images of galaxy clusters of masses
    $\Mf\approx 3 \times 10^{14}\Msol$ (left), $\Mf\approx
    5 \times 10^{14}\Msol$ (centre), and $\Mf\approx
    11 \times 10^{14}\Msol$ (right). The images are scaled so
    that each is $2.4\Rf$ on a side. This Figure is adapted from \citet{che21}. }
\label{fig:chex-mate}
\end{figure}

The self-similar appearance of galaxy clusters is illustrated in Figure \ref{fig:chex-mate}, which shows X-ray images of clusters observed with \XMM. 
The clusters span a range in mass of $\Mf\approx$ 3--11$\times 10^{14}\Msol$, but when the images are scaled in size relative to $\Rf$ for each cluster, their overall appearance is very similar
(note that the differences in X-ray emission away from the centres of the clusters are due to different levels of residual background emission).

Under the assumptions above, the properties of galaxy clusters measured within overdensity radii are expected to scale with mass following simple power-law forms. 
Here we will focus on these scaling relations for the principal properties of the ICM that can be measured with X-ray observations; namely its mass ($\Mgas$), temperature ($\Tx$), and X-ray luminosity ($\Lx$), along with quantities derived from combinations of those properties. 
Similar relations can also be derived for other properties (such as the total stellar mass, or the velocity dispersion of the member galaxies).

We note that the use of overdensity radii in the self-similar model means that clusters of a given mass are denser at higher redshift, having formed in a denser Universe (see Equation (\ref{eq:Rdelta})). 
This leads to the expectation of certain cluster properties evolving, with a dependence on $E(z)$. 
So, for example, clusters of a given mass would be hotter and more luminous at higher redshifts. 
This is a simple consequence of defining the mass in terms of an overdensity, and does not describe any astrophysical evolution of galaxy clusters; rather it is the baseline against which astrophysical evolution should be measured.

In the following, all properties are measured within an overdensity radius, but we drop the $\Delta$ subscript for clarity. 
While deriving the self-similar behaviour for the scaling relations, we will also generalise them in the form
\begin{align}\label{eq:slfit}
Y = A E(z)^\gamma X^B
\end{align}
where $A$, $B$, and $\gamma$ describe the normalisation, slope, and evolution respectively of the scaling relation between some properties $X$ and $Y$. 
The values of $B$ and $\gamma$ predicted by the self-similar model for each of the scaling relations considered are summarised in Table \ref{table:self}.

We note that $E(z)^\gamma$ is not a unique way to parametrise the evolution. 
A common alternative is to fix $\gamma$ to the self-similar value and then parametrise additional evolution as a power of $1+z$.
Either variation only describes evolution in the normalisation of the scaling relations, and extensions include allowing for a redshift dependence of the slope and scatter of the scaling relations.

\subsubsection{The \MgM\ relation}
\label{sec:mgm-relation}
For self-similar clusters, the mass of the ICM, $\Mgas$, is a constant fraction, $\fgas$, of the total mass of the cluster:
\begin{align}
  \label{eq:mgm1}
  \Mgas = \fgas M.
\end{align}
This can be generalised as
\begin{align}
\Mgas = \Amgm E(z)^{\gmgm} M^{\Bmgm}
\end{align}
where for self-similarity $\gmgm=0$ and $\Bmgm=1$.

\subsubsection{The \TM\ relation}
\label{sec:mt-relation}
If the ICM is isothermal with a temperature $\Tx$, and its particles are in virial equilibrium with the cluster potential, then the virial theorem gives
\begin{align}
\Tx \propto \frac{M}{R} .
\end{align}
Eliminating $R$ in favour of $M$ using Equation (\ref{eq:Rdelta}), and noting that $\rho_c(z)\propto E(z)^2$, we can write the self-similar
\TM\ relation:
\begin{align}
  \label{eq:tm1}
  \Tx \propto E(z)^{2/3}M^{2/3}.
\end{align}
This generalises to
\begin{align}
\Tx = \Atm E(z)^{\gtm} M^{\Btm}
\end{align}
with $\gtm=2/3$ and $\Btm=2/3$ for self-similarity.

\subsubsection{The \LM\ relation}
\label{sec:lm-relation}
In order to derive the self-similar form of the scaling with mass of the X-ray luminosity of the ICM ($\Lx$), we will make the approximation that the X-ray emission is due solely to the bremsstrahlung mechanism. 
Neglecting the weak temperature dependence of the Gaunt factor in the bremsstrahlung emissivity (\citealt{boh10}), the bolometric bremsstrahlung luminosity of the ICM is given by
\begin{align}
  \label{eq:brem}
  \Lbol \propto \int \epsilon dV \propto \int \rhog^2 \Tx^{1/2} dV,
\end{align}
where $\rhog$ is the density of the ICM, and $V$ is the volume of the cluster. The last scaling is strictly valid only for massive systems for which the cooling function mainly depends only on \Tx~(i.e., $\Lambda(\Tx,Z)\sim\Tx^{1/2}$).

The integral in Equation (\ref{eq:brem}) depends on the temperature and density distribution of the ICM. However, for self-similar clusters the ICM distributions are identical, and assuming isothermality, $\Tx$ is a constant. This leads to
\begin{align}
  \label{eq:l-rho}
  \Lbol \propto \rhog^2 \Tx^{1/2} R^3.
\end{align}
For self-similar clusters, $\rhog\propto M/R^3$, and we can use
Equation (\ref{eq:Rdelta}) again to eliminate $R$ in
favour of $M$ and $E(z)$ to give
\begin{align}
\Lbol \propto E(z)^2 M \Tx^{1/2}.
\end{align}
Finally, we can use Equation (\ref{eq:tm1}) to eliminate $\Tx$ giving the
self-similar bolometric \LM\ relation
\begin{align}
  \label{eq:lm1}
  \Lbol \propto E(z)^{7/3} M^{4/3}.
\end{align}

The \LM\ relation may then be generalised as
\begin{align}
\Lx = \Alm E(z)^{\glm} M^{\Blm}
\end{align}
with $\glm=7/3$ and $\Blm=4/3$ for self-similar behaviour and
bolometric luminosity.

For soft-band luminosities, the temperature dependence of the bremsstrahlung emissivity is very small, and can be neglected in Equation (\ref{eq:brem}). This leads to the self-similar soft band
\LM\ relation
\begin{align}
\label{eq:lmsoft}
  \Lsoft \propto E(z)^{2} M
\end{align}
i.e., $\glm=2$ and $\Blm=1$ for soft-band luminosities.

In the preceding derivation, we assumed that the X-ray emission from the ICM was entirely bremsstrahlung. 
However, as we saw in \textsection \ref{sec:x-ray-emission}, line emission makes an increasingly important contribution to the emissivity at lower temperatures. 
Taking this into account, the temperature dependence of the bolometric X-ray luminosity changes from $\Tx^{1/2}$ in the bremsstrahlung-dominated regime (as in Equation (\ref{eq:brem})) towards approximately $\Tx^{-1/2}$ for $\Tx\lesssim 2\keV$ (see \citealt{lov21} for a more detailed discussion). 
For the lowest temperature clusters, the self-similar slope and evolution of the bolometric \LM\ relation then tend towards $\Blm=2/3$ and $\glm=5/3$.

\subsubsection{The \YM\ relation}
\label{sec:ym-relation}
Since $\Tx$ is proportional to the kinetic energy of an ICM particle, and $\Mgas$ is proportional to the number of particles in the ICM, their product $\Yx\equiv\Tx\Mgas$ is proportional to the total thermal energy of the ICM (\citealt{2006ApJ...650..128K}). 
In the self-similar model, this energy derives from the gravitational collapse of the ICM so should correlate directly with the total cluster mass.

The self-similar form of the \YM\ relation is easy to derive by
combining Equations (\ref{eq:mgm1}) and (\ref{eq:tm1}) to give
\begin{align}
  \label{eq:ym1}
  \Yx \propto E(z)^{2/3}M^{5/3}
\end{align}
or
\begin{align}
  \label{eq:ym2}
\Yx = \Aym E(z)^{\gym} M^{\Bym}
\end{align}
with $\gym=2/3$ and $\Bym=5/3$ for self-similarity.

\subsubsection{The \LT\ relation}
\label{sec:lt-relation}
Scaling relations between different combinations of ICM properties can be derived by combining the relations above to eliminate mass. This is often desirable since it allows the scaling behaviour of clusters to be studied using more directly observable quantities. The most important of these is the scaling relation between $\Lx$ and $\Tx$.

Combining Equations (\ref{eq:tm1}) and (\ref{eq:lm1}), we can derive the self-similar form of the bolometric \LT\ relation:
\begin{align}
  \label{eq:lt1}
  \Lx \propto E(z) \Tx^2.
\end{align}
Once more we can generalise this as
\begin{align}
  \label{eq:lt2}
\Lx = \Alt E(z)^{\glt} \Tx^{\Blt}
\end{align}
with $\glt=1$ and $\Blt=2$ for self-similar scaling.

As is the case with the \LM\ relation, the expected self-similar slope and evolution of the \LT\ relation will depend on the energy band in which the luminosity is measured. 
Furthermore, the increasing contribution of line emission to the ICM luminosity in lower temperature clusters will also alter the self-similar slope and evolution for lower temperature clusters. These effects can be easily incorporated into the \LT\ relation by following the approach adopted for the \LM\ relation.

\begin{table}[t!]
\caption{Self-similar values for slope $B$ and redshift evolution $\gamma$ of the scaling relations. As discussed in the text, the slope values for the luminosity scaling relations are appropriate for hotter systems where the X-ray emission is dominated by bremsstrahlung.}
\centering
\begin{tabular}{m{3cm} m{2cm} m{0.5cm}}
\toprule
Relation (Y, X) & $B_\text{self}$ & $\gamma_\text{self}$\\
\midrule
\Mgas--\Mf & 1 & 0 \\
\Tx--\Mf & 2/3 & 2/3 \\
\Lbol--\Mf & 4/3 & 7/3 \\
\Lsoft--\Mf & 1 & 2 \\
\Lbol--\Tx & 2 & 1 \\
\Lsoft--\Tx & 3/2 & 1 \\
\Yx--\Mf & 5/3 & 2/3 \\
$K$--\Tx & 1 & -4/3 \\
\bottomrule
\label{table:self}
\end{tabular}
\end{table}

\subsubsection{The entropy of the ICM}
\label{sec:entropy-icm}

The entropy of the ICM is a useful quantity with which to examine the self-similarity (or lack thereof) of clusters (see \citealt{voi05a}, for a more detailed introduction to the topic). It is conventional to define entropy of the ICM as
\begin{align}
  \label{eq:entropy}
  K = \Tx \nel^{-2/3}
\end{align}
which is proportional to the logarithm of the traditional thermodynamic entropy. This quantity can be quite easily measured with X-ray observations.  For self-similar clusters the density is proportional to $E(z)^2$ and therefore is straightforward to derive  the K--\Tx\ relation:
\begin{align}
  \label{eq:kt1}
  K \propto E(z)^{-4/3} \Tx.
\end{align}
Once more we can generalise this as
\begin{align}
  \label{eq:kt2}
K = \Akt E(z)^{\gkt} \Tx^{\Bkt}
\end{align}
with $\gkt=-4/3$ and $\Bkt=1$ for self-similar scaling.

Entropy is not changed by adiabatic processes, so has simple behaviour in the self-similar model, and tracks processes that cause departures from self-similarity. 
Low entropy gas will sink in a gravitational potential, while high entropy gas rises buoyantly. 
The entropy of the ICM in an undisturbed cluster will thus simply increase monotonically with radius.

For self-similar clusters, in which the gas that collapses to form the ICM is cold, the ICM gains entropy at the shock that is formed where the accreting gas meets the already virialised ICM. 
The strength of the shock depends on the mass of the cluster, so layers of ICM with increasing entropy are simply added as the cluster grows. 
This leads to a simple radial power law of the form $K\propto r^{1.1}$ for smooth, spherical accretion (see, e.g., \citealt{2001ApJ...546...63T}).

If the infalling gas has been ``pre-heated'' (i.e., has some thermal energy and hence non-zero entropy prior to accretion), then the accretion shock is weaker (since the sound speed in the inflowing gas is higher), and so less entropy is added by the shock. 
The net effect is that the entropy of the shocked gas is the same as it would have been without pre-heating, but there is now a minimum entropy level (the amount provided by the pre-heating), often referred as an ``entropy floor''. 
Thus, the entropy profile of the ICM will flatten towards the minimum level in the central parts of the cluster instead of continuing a power-law form towards $r=0$.

Any energy input in the ICM (either pre-heating or once the gas has
accreted) will raise the entropy of the affected gas which will
then change its location in the cluster. 
Pre-heated gas will only be able to sink as far into the cluster as its entropy permits.
Similarly, gas whose entropy is increased within the cluster will rise through convection until it reaches a radius with a matching entropy.
If the entropy were raised enough, then the gas would be unable to sink down to (or would rise beyond) the inner regions of clusters where the ICM is bright enough to be observed in X-rays. 
This would impact ICM scaling relations by reducing the mass of ICM observed within a cluster, and hence also its X-ray luminosity.

The effects of in-situ heating or preheating should also be observable in radial profiles of thermodynamic properties of the ICM, such as entropy. However, these can be difficult to measure unambiguously due to effects such as the clumping of ICM biasing measurements of its density (see Kay et al. - Chapter: {\it Thermodynamical profiles of clusters and groups, and their evolution}, of this volume).

\subsection{Heating and cooling the ICM}
\label{sec:non-grav-proc}

The self-similar relations derived in the previous sections assume that gravitational heating dominates in clusters. 
The basic idea is that due to the deep gravitational potential wells the gas falling into a cluster (either at the same time as the dark matter, or subsequently) will cause it to move very rapidly, colliding with other gas, and being shocked. 
However, in reality things are much more complicated and there are a number of heating and cooling processes at work in galaxy clusters.
We will briefly introduce the relevant processes here, and return to them in more detail when interpreting the observations of scaling relations in \textsection \ref{sec:interpretation}.

In contrast to the self-similar model, real clusters do not form in a single spherical collapse. Instead they form hierarchically through episodic mergers of smaller mass structures to build larger systems.
Clusters grow continuously through the accretion of smaller structures punctuated by occasional major mergers with other objects of a similar mass. The shocks produced by major mergers dissipate kinetic energies of 10$^{63}$--10$^{64}$ ergs, and while the ICM will ultimately approach virial equilibrium with the new gravitational potential (although it may take a few Gyr to reach equilibrium again; e.g., \citealt{2012ApJ...751..121N}), its
temperature and luminosity, and density structure will all be
perturbed significantly during the course of the merger. This would
produce offsets from the mean scaling relations that may change
significantly at different points during the merger.

The emission of X-rays is a strong cooling mechanism for the ICM. The bolometric emissivity due to bremsstrahlung is proportional to $\nel^2\Tx^{1/2}$, and in the dense central regions of clusters, the cooling time of the ICM may be short compared to the Hubble time. 
At high temperatures (where thermal bremsstrahlung dominates), the
cooling time scales as
\begin{equation}
t_{cool}\propto \Tx^{1/2} \nel^{-1}.
\end{equation}
At constant pressure the rise in density as the temperature drops means that cooling accelerates as the gas cools. This leads to the formation of {\it cool cores}: very bright, sharply peaked regions of X-ray emission, associated with cooler gas in the central $\lesssim 100$~kpc of a cluster. Conversely, clusters with no central drop of temperature and flatter surface brightness profile (i.e., with high central entropies) are called {\it non-cool cores}.

The X-ray luminosity of a cool core can account for more than $50\%$ of the total X-ray emission from a cluster, but not all clusters possesses cool cores (likely due to them being disrupted by significant mergers). 
This gives rise to a very significant scatter in X-ray luminosity, which can be significantly reduced if the core regions (conventionally within a radius of $0.15\Rf$) are excluded. 
This is not always possible, for example if the observation lacks the angular resolution to resolve the core, or the data quality does not permit the exclusion of a significant fraction of the photons.

In principle this cooling is a runaway process, with gas condensing out of the X-ray emitting regime in the cores of clusters, and being replaced by a steady flow of gas cooling from larger radii. However, high resolution observations of cluster cores have not found the expected amounts of cool gas (e.g., \citealt{2001A&A...365L.104P}). It is now clear that some form of energy input regulates the cooling, allowing the formation of stable cool cores.

Active galactic nuclei (AGNs) at the center of galaxy clusters have emerged as heating agents powerful enough to prevent further cooling.
In support of this scenario, there are many observations of galaxy clusters with striking cavities in their X-ray emission, which have been interpreted as bubbles of relativistic plasma inflated by radio jets (see \citealt{2021Univ....7..142E} and references therein).
Estimates of the energetics of these bubbles show that there is
sufficient power output from the AGN to counterbalance the radiative cooling\footnote{In principle, thermal conduction can transport energy down the temperature gradients into cool cores (e.g.,  \citealt{2001ApJ...562L.129N}), but it is likely to only increase the central radiative cooling time by a factor of a few, failing to prevent runaway cooling (e.g., \citealt{2005MNRAS.364...13P}).}.

The effects of AGN feedback are expected to be two-fold. In addition to the ongoing energy input through shocks and bubbles that prevents the runaway cooling of the ICM in cluster cores, AGN also input significant energy to the ICM via radiation at high redshift \citep{fab12}. 
This ``radiative'' mode of AGN feedback is thought to have raised the entropy and hence reduced the gas content of low-mass clusters (whose shallow gravitational potential cannot retain the heated gas). 
This would then be apparent in departures from self-similar scaling of ICM properties (see \textsection \ref{sec:interpretation}).

Feedback from supernovae and galaxy winds can operate in a similar way to AGN feedback  (e.g., \citealt{2000MNRAS.311...50M}), although these may not be energetic enough to have a significant effect above the scales of low mass groups.

The processes described above (i.e., cooling, AGN and SNe feedback, merging) are often referred to as ``non-gravitational'' heating processes, and play an important role in determining scatter, shape, and evolution of the scaling relations. Therefore, they can provide information on the physical mechanisms causing the departures from the self-similar predictions.

\section{Analysis methods and considerations}
Slopes, normalizations, and scatters of cluster scaling relations can vary notably as a result of different selection functions, observational biases, observational uncertainties, systematic errors from heterogeneous data reduction and analysis methods, band-pass sensitivities, and other effects. Removing all biases and systematic differences is very challenging, but the correct astrophysical interpretation of the scaling relations, and their reliable use for cosmological studies, requires that these systematics are accounted for. In the following we provide a brief summary of the main issues that should be considered when modelling the cluster scaling relations.

\subsection{Observational biases}
\label{sec:observational-biases}
The starting point for any analysis is the definition of the sample to be used. In astrophysical scenarios, samples are rarely perfectly representative of the parent population, leading to selection biases.  These must be modelled in order to recover the properties of the true population.

\begin{figure}[t]
  \centering
  \includegraphics[width =0.95\textwidth]{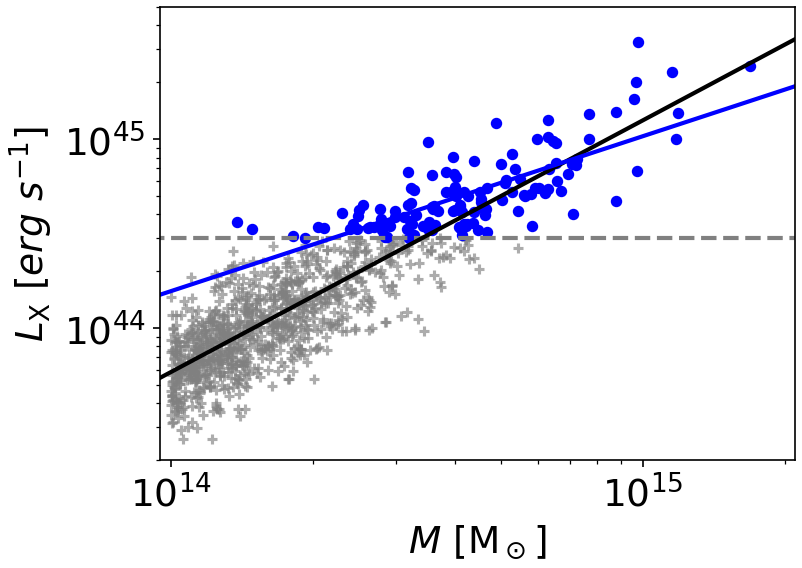}
  \caption{Illustration showing the effect of Eddington bias on
    scaling relation data. An example \LM\ relation was simulated by
    drawing masses from a model mass function (resulting in the much
    larger numbers of points with lower masses in the plot), and then
    assigning a luminosity to each point assuming a reference \LM\
    relation with $50\%$ intrinsic scatter. This input scaling
    relation is shown by the solid black line. A simple luminosity
    selection was applied to produce a mock observed sample, indicated
    by the grey dashed line. The blue circles show the clusters that
    were included in the sample and the grey crosses denote the
    clusters that were excluded. The solid blue line is a simple fit
    to the mock observed sample, and is clearly strongly biased
    relative to the relation from which the sample was drawn.}
  \label{fig:biases}
\end{figure}

As a simplistic example, we will consider the measurement of the \LM\ relation for a sample defined by selecting clusters in an X-ray survey that have measured luminosities above some threshold (real samples will have more complicated selection functions, primarily dependent on observed quantities like X-ray flux, as discussed below). If there is some degree of scatter in the luminosity of clusters of the same mass (due to intrinsic scatter\footnote{The intrinsic scatter is the dispersion around the ideal straight line distribution which would manifest even without measurement error in x and y. In other words, the intrinsic scatter is the difference between the measured variance around the best-fit and the variance due to measurement errors.} or measurement noise) then for clusters close to the luminosity threshold, those with below-average luminosities are more likely to be excluded from the sample. The result is shown in Fig. \ref{fig:biases}, where the distribution of observed clusters (blue points) is obviously not following the underlying relation indicated by the black line. If one does not account for this effect when fitting the observed data point, the fitted relation (blue line) will be flatter.

This type of selection bias will be present in many areas of statistical analysis, but there are some additional complications that are often important in astrophysical scenarios.

\begin{figure}[t]
  \centering
  \includegraphics[width =0.95\textwidth]{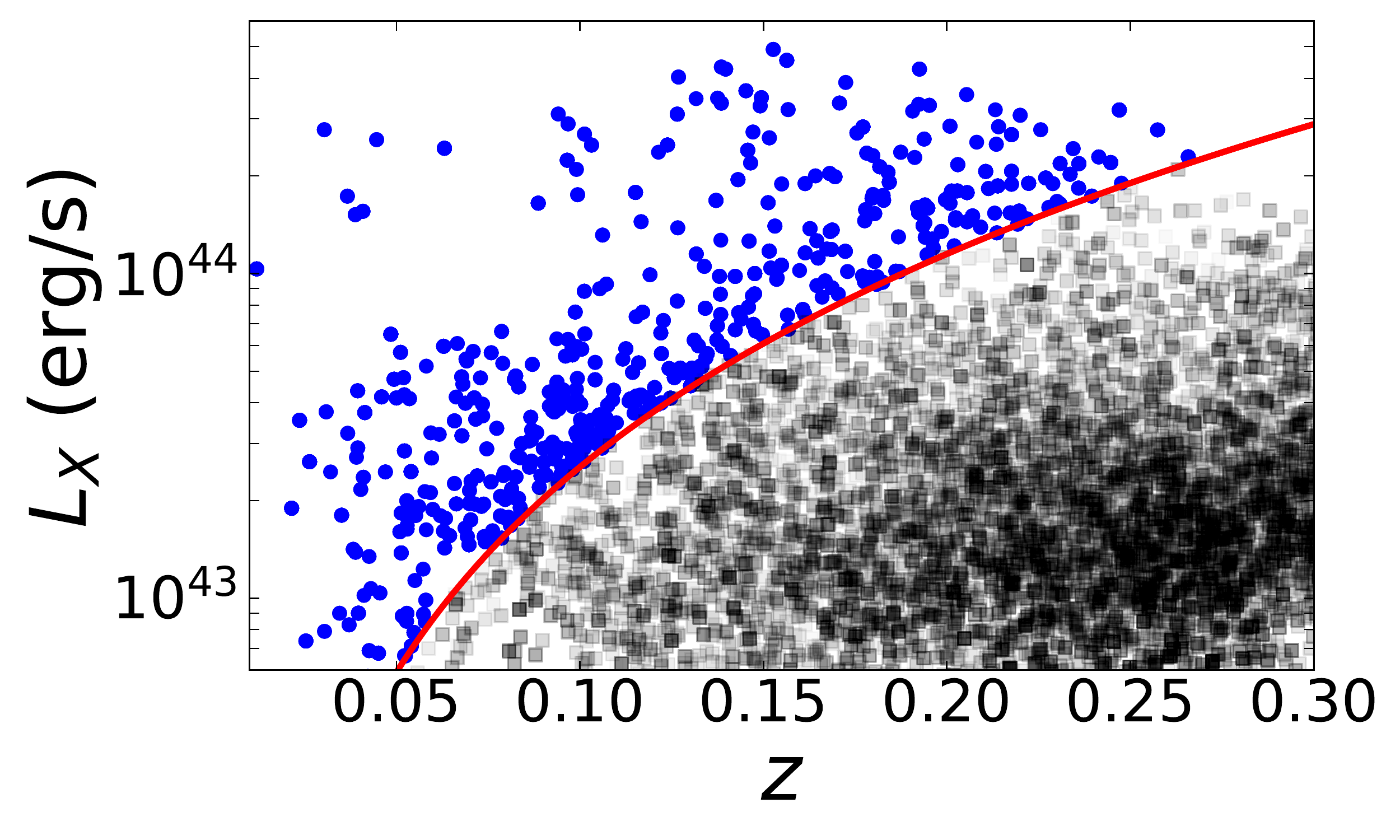}
  \caption{Illustration showing the effect of the Malmquist bias on a flux-limited sample. A sample of clusters was simulated by drawing masses and redshifts from a model mass function for a hypothetical area of sky, and luminosities were assigned to each cluster assuming a simple scaling relation with mass. The data points show the clusters plotted on the luminosity-redshift plane. The full sample is dominated by distant, low luminosity (low mass) clusters. The blue circles denote the clusters that would be included in an example flux-limited survey (with the flux limit illustrated by the red line).}
  \label{fig:malmquist}
\end{figure}

The most famous is probably the {\it Malmquist bias} (\citealt{1922MeLuF.100....1M}), an effect caused by the fact that brighter sources are detectable at greater distances. 
This means that clusters with luminosities that are above average for their mass are detectable over a larger volume of the Universe than those with luminosities below average for their mass. 
The brighter clusters will thus be over-represented in a sample selected by X-ray flux. This is illustrated in Figure \ref{fig:malmquist}, which shows how clusters with a higher luminosity are more likely to be found in a flux-limited sample.

This picture is further complicated by the presence of the {\it Eddington bias} (\citealt{1913MNRAS..73..359E,1940MNRAS.100..354E}), which arises when a sample is defined by a threshold in some property which has a noisy correlation with the property of interest. 
Suppose a sample of clusters is selected on the basis of their luminosity or flux. If there is scatter in luminosity at a given mass (due to intrinsic scatter or measurement noise), there will be some clusters whose average luminosity for their mass is below the selection threshold, but their scatter in luminosity carries them above the threshold. 
Similarly there will be clusters whose average luminosity for their mass is above the threshold but are lost from the sample as their scatter in luminosity carries them below the threshold. 
This biases the sample towards clusters that have above-average luminosities. 
Furthermore, the number of clusters as a function of their mass (the mass function) decreases rapidly with increasing mass. 
This means that there are more low mass clusters available to be scattered into the sample than high mass clusters to be scattered out of the sample, increasing the magnitude of the bias. This is illustrated in Figure \ref{fig:biases}.

The terms Malquist and Eddington bias are sometimes used interchangeably and conflated in the literature, and indeed their effects are similar and will both be present in a sample selected by flux.

These selection biases impact most directly on the property used to define the sample (luminosity in the above examples). 
However, any other cluster properties that have covariance with the property used for selection will also be impacted. For example, we expect from Equation \eqref{eq:brem} that clusters with above-average luminosity relative to their mass will also tend to have above-average \Mgas. 
A luminosity-based selection would thus confer some degree of bias on scaling relations involving \Mgas.

On top of the {\it Malmquist} and {\it Eddington biases}, there is a {\it morphological bias}.
As we will see in Section \ref{sec:obs-scaling}, relaxed and disturbed clusters populate different regions of the residual space with respect to the best-fit of the luminosity-mass and luminosity-temperature relations, with the relaxed objects on average above the relations and disturbed systems on average below (e.g., \citealt{pra09a}, \citealt{lov20}).
This offset is probably associated with the strength of cool-cores that boost the X-ray luminosity. Mergers also contribute to the scatter because the total masses can easily be incorrectly estimated when the clusters are not in HE as happens during cluster mergers, which can change the X-ray luminosities and temperatures of clusters, both in a transient and in a long term manner.
Thus, if the sample definition depends directly or indirectly on the morphology of clusters, this could lead to observed relations that are different in slope, normalization, and scatter for samples with different selection criteria.
For instance, X-ray selected samples are known to be significantly biased towards cool-core clusters with a larger effect for low-mass systems (\citealt{2011A&A...526A..79E}).

Corrections of the above-described biases can be applied if the selection function is known. 
This is clearly not the case for samples constructed from public archives which are usually over-represented by clusters with peculiar morphology or special properties (e.g., strong cool core clusters or major mergers) than more ``ordinary'' clusters.
This is known as {\it archival bias}.

The magnitude of these biases differs from sample to sample, and depends on the amount of true scatter (i.e., likely greater than that observed) in the relations, with larger values requiring a larger correction.
Moreover, it also depends on the slope of the mass function around the detection threshold of the sample. Because of the steepening of the mass function at high masses, surveys with higher mass limits will suffer a larger bias. 
The situation can be further complicated by a non-uniform magnitude of the scatter. 
In fact, the intrinsic scatter of galaxy groups may be higher than that of clusters due to the larger impact of the non-gravitational processes on their global X-ray properties (e.g., \citealt{lov21} and references therein).

We will mention a final type of bias that is independent from the sample definition, and is often underestimated, the so-called {\it confirmation bias}.
This is a type of cognitive bias that involves favoring information that confirms previously existing beliefs.
Confirmation biases impact how information is collected, and how the results are interpreted.
In fact, researchers may tend to select information that supports their original views, ignoring inconsistent results, or interpret uncertain evidence as supporting their original ideas. An example, when studying scaling relations, is to perform a new (more accurate) analysis on systems that fall far away from the
expected predictions under the (often wrong) assumption that something went wrong in the analysis.
This could, for example, artificially reduce the scatter and change the slope of the relation. Confirmation bias can be avoided through the use of blinding in analyses, so that researchers do not view the final results of the analysis until the methodology is tested and frozen. This approach is starting to be adopted in astrophysics, but can be very difficult to implement in practice.

\subsection{Selection effects and selection functions}\label{selection}
In order to account for the selection biases discussed above, and derive the true scaling relations of the underlying cluster population, as opposed to the merely observational trends in the studied sample, one needs to understand the sample definition. This requires the determination of the survey selection function which describes the probability of detecting an object given a set of cluster properties (e.g., redshift, luminosity, total mass, dynamical state), survey and instrumental conditions (e.g., sensitivity, spatial and/or spectral resolutions), and detection algorithm used (e.g., sliding cell or wavelet methods in X-ray imaging surveys).

The derivation of a selection function for X-ray galaxy clusters can be split into two parts: the detection of X-ray sources and their classification as clusters.
X-ray surveys are dominated by unresolved point sources (primarily AGNs), which outnumber galaxy clusters by of order $100:1$. Sources are primarily classified as clusters on the basis of being extended (i.e., resolved) objects.

Ideally one would like to obtain a sample which is complete (i.e., all objects fulfilling certain observational criteria are detected), and uncontaminated (i.e., all objects not fulfilling such observational criteria are excluded). The detection process for clusters in X-ray images is complicated by their generally low surface brightness. This reduces their contrast with the background and results in sensitivity on the background conditions and modelling.
This may lead to clusters with irregular and diffuse morphologies being missed by a survey, while a more regular cluster that is centrally peaked due to a cool core would be easier to detect.
Moreover, one needs to deal with spurious (false) detections due to noise fluctuations or instrumental effects, and the addition of misclassified point sources to the cluster sample.
Finally, point sources can overlap the emission from a cluster (due to an AGN in the cluster or a chance projection along the line of sight). Depending on the flux and location of the point source, it may lead to the cluster being misclassified as an AGN, or it could contribute to the total flux of the cluster (e.g., \citealt{2018MNRAS.481.2213B}).

Understanding the selection function of any survey is not an easy task because, by its very nature, the selection function is supposed to tell us about objects that have not been detected. 
This is often achieved by running the survey's detection algorithm on mock observations obtained from realistic simulations. 
Since one knows which objects exist in the simulations, this procedure allows the cluster detection probability to be determined as a function of some basic input parameters (e.g., fluxes and sizes) and accounting for a range of instrumental (e.g., PSF, vignetting) and astrophysical (e.g., different background conditions) effects.

Since the selection effects are specific to a particular catalogue extraction method, by comparing different codes and cluster detection methods one can better understand the selection effects and catalogue uncertainties, and their relation with the cluster properties. 
In practice, as highlighted by \cite{2005A&A...429..417M}, a complete understanding of a selection function can only be obtained by combining such simulations with observations taken under different conditions and/or in different wavelengths. 
Thus, our understanding of the selection function, and therefore of the underlying cluster population, is a slow process.

\subsection{X-ray vs Optically and SZ selected samples}\label{xoptsz}

Historically, optical and X-ray surveys have provided the primary source of cluster catalogues.
However, in the recent years galaxy clusters have been also detected through the Sunyaev-Zel'dovich (SZ) effect, the spectral distortion of the cosmic microwave background (CMB) generated via inverse-Compton scattering of CMB photons by the hot electrons in the ICM (\citealt{1972CoASP...4..173S}).
The strength of the SZ signal is independent of the redshift of the scattering cluster and is tightly correlated with the cluster's total mass.
This means SZ-selected samples are as near as possible to being mass-selected, and as such are considered to be close to unbiased.
Comparisons between properties of samples selected with these different approaches are giving insight into the impact of the selection effects on the derived cluster properties.

While the ICM is responsible for both the X-ray light and SZ, the observables have a different dependence on its distribution with a cluster.
The X-ray emission scales with the square of the ICM density while the SZ scales linearly. This means that X-ray surveys are expected to detect a larger fraction of centrally peaked and relaxed clusters than SZ experiments.
Indeed, this is confirmed by several observational studies (e.g., \citealt{2016MNRAS.457.4515R,2017MNRAS.468.1917R},
\citealt{2017ApJ...843...76A}, \citealt{2017ApJ...846...51L}; but see \citealt{2020MNRAS.495..705Z} for a different view).

That said, both the X-ray and SZ signals will depend on the overall amount of ICM in a cluster, and so may share biases towards detecting clusters with higher gas fractions (e.g.,
\citealt{2016A&A...585A.147A}). 
The impact is likely to be strongest for low mass clusters where the gas fraction is significantly lower than that of massive clusters (e.g., \citealt{2015A&A...573A.118L}, \citealt{2016A&A...592A..12E}), making the selection bias mass-dependent.

One possible solution is to use optical selection to identify groups and clusters, which can be then observed with X-ray/SZ telescopes. 
The ICM properties of these optically-selected clusters can then be compared with those of X-ray and SZ selected samples. 
A limitation of this approach is that current optical selections suffer from projection effects (e.g., \citealt{2019MNRAS.482..490C} and references therein) which lead to contaminated lists of cluster
candidates. 
Another independent selection can be achieved through the weak-lensing signature of clusters, although the lensing signal can be affected by the projection of foreground and background structures along the line-of-sight (see \citealt{2018PASJ...70S..27M} and references therein).

Clusters selected in the optical via overdensities of galaxies or via their weak lensing signal tend to be X-ray underluminous for a given mass by a factor of $\sim$2-3 compared with X-ray and SZ selected clusters (e.g., \citealt{2015MNRAS.447.3044G},
\citealt{2016A&A...585A.147A}).
This may be due to the ICM-based detection methods being biased against low \fgas\ (and hence low gas luminosity and pressure) clusters.
However, further investigations are needed to understand whether these clusters are really X-ray underluminous or their masses are overestimated (e.g., due to projection effects).

Thanks to the large multiwavelength data-sets which are becoming
available, we are now in the position to compare the cluster
detections based on multiple independent observables in order to
cross-validate survey selection functions.

\subsection{Correlated errors}
When modelling the correlations between cluster properties, it is of course necessary to include the measurement errors on the quantities.
One must also consider whether the errors on the quantities of
interest are independent or correlated.
Suppose an observation were repeated a large number of times, and the same two properties were measured each time, giving a distribution of values of each.
If the errors were correlated, then the distributions would be correlated (i.e., an above-average value of one quantity would correspond to a below average value of the second quantity in the case of a negative correlation).

While independent measurement errors are usually described with a one-dimensional normal distribution when modelling data, correlated errors require a multivariate normal distribution (i.e., with one dimension per property being measured) parameterised by a covariance matrix.
The covariance matrix describes the size of the measurement errors and the degree of correlation between them.
This introduces a degree of complexity to the analysis, and has been neglected in most analyses of the X-ray scaling relations, but may be essential if the measurement errors and their correlations are not negligible.

A degree of correlation is expected between the measurements of all of the X-ray properties derived from a given observation, because the photons used for the estimation are the same. 
However, depending on the analysis method this correlation can be reduced. 
For instance, if the luminosity and temperature measurements are both based on the spectroscopic analysis (for massive clusters the luminosity is given by normalization of the spectrum, while the temperature is determined from its shape), the correlation of their measurement errors is high.
If a surface brightness profile is used to estimate the luminosity
instead, then the measurements are nearly uncorrelated.

The correlations between measurements can be complex. 
Since HE mass measurements are based on gas density and temperature profiles, correlated measurement errors exist between \Mf\ and both \Mgas\ and \Tx. 
Furthermore, one may then measure the value of cluster properties within a radius defined from the hydrostatic mass measurement (e.g., \Rf), introducing additional correlations in the measurements (in particular for \Mgas\ and \Lx). 
To minimize the correlation between measurements one can have an independent estimate of the region within which determining the X-ray properties (e.g., \citealt{2020MNRAS.492.4528S}), but this comes at the cost of an increased intrinsic scatter. 
In general, the correlation between the errors can be measured through Monte Carlo simulations and then incorporated into the analysis of the scaling relations (e.g., \citealt{man16}).

\subsection{Linear regression and fitting packages}
Cluster properties follow power-law correlations, so are linear in log space and hence the measurement of the scaling relation parameters generally involves some form of linear regression, typically in the form of Eq. \ref{eq:slfit} where $X$ and $Y$ are usually renormalized by values, $X_0$ and $Y_0$, chosen to bring the range of $X$ and $Y$ close to unity, which is beneficial for some modelling approaches. 
Setting $X_0$ and $Y_0$ close to the median value of $X$ and $Y$ for the data being modelled will also reduce the degeneracy between $A$ and $B$ in the model.

Linear regression is one of the most widely used statistical techniques in science, but there are numerous complicating factors in the case of astronomical data.
These may include: (i) non-negligible, heteroscedastic, measurement errors; (ii) correlations between the measurement errors on the quantities being compared; (iii) no meaningful way to define a dependent or independent variable; (iv) significant intrinsic scatter of the data; (v) a non-uniform distribution of data (e.g., many more low-mass clusters than high-mass clusters); and (vi) missing data due to sample selection (see Sect. \ref{selection}).
These aspects can have a non-negligible effect on the regression results if not properly accounted for (see, e.g., \citealt{1996ApJ...470..706A} for the impact of measurement errors and intrinsic scatter, and \citealt{2007ApJ...665.1489K} for the effect of non-uniformly distributed variables).

\begin{figure}[t]
  \centering
  \includegraphics[width =\textwidth]{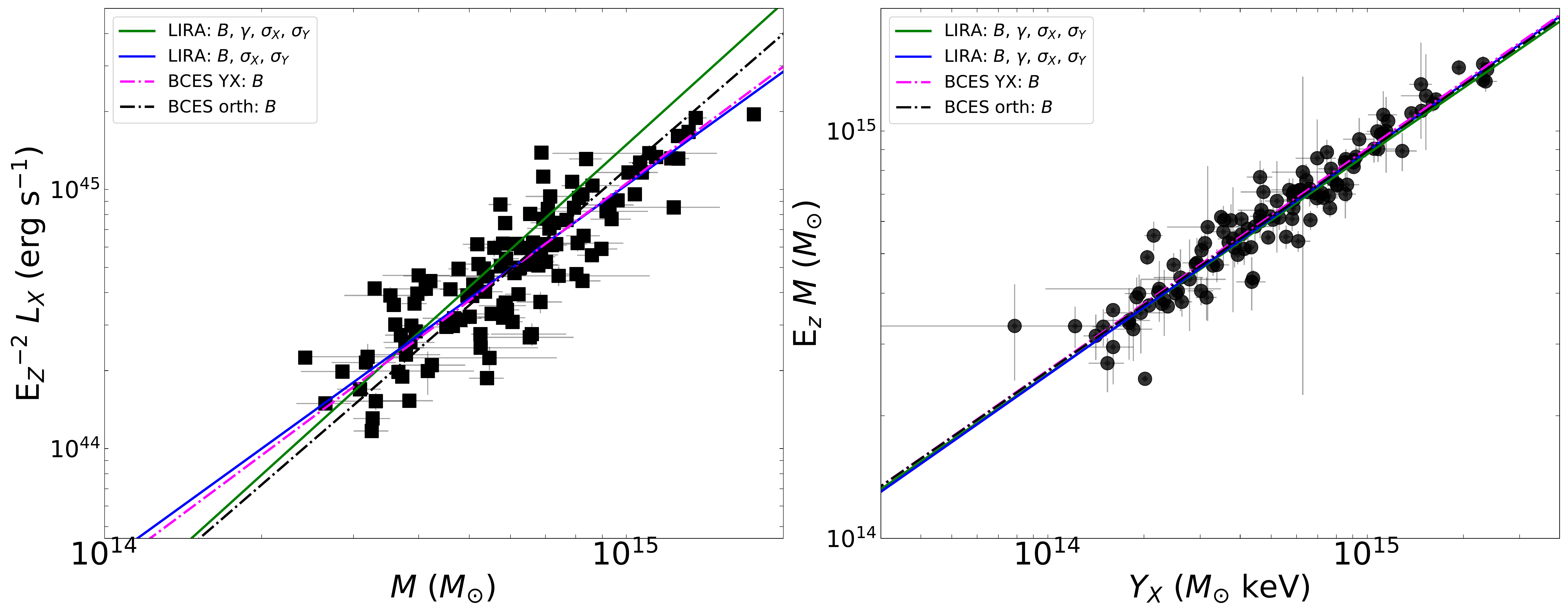}
  \caption{Illustration adapted from \citet{lov20} showing the comparison of the fitted results for the \LM\ ({\it left})  and \MY\ ({\it right}) relations obtained using different fitting methods. In green we show the fitted relation obtained using LIRA (\citealt{2016MNRAS.455.2149S}) allowing all the parameters to vary (i.e., slope, normalization, evolution, and both scatters). In blue we show the best-fit results obtained assuming the self-similar redshift evolution.  In black and magenta we show the result from BCES (\citealt{1996ApJ...470..706A}), where the redshift evolution was chosen to be self-similar.}
  \label{fig:fitpack}
\end{figure}

Different analyses of scaling relations have used different approaches to address some or all of the issues above, complicating the comparison and the interpretation of the results (as highlighted in \citealt{lov20}). This is illustrated in Figure \ref{fig:fitpack}.
\cite{2013SADM....6...15A} provide an overview of common methods for fitting scaling relations, discussing some pros and cons for each technique.
Broadly speaking, the different approaches can be divided into {\it  frequentist} and {\it Bayesian}.
Both methods rely on computing the \textit{likelihood} (the probability of obtaining the data that was observed), given a model and set of parameter values.
In the frequentist approach, the probability distributions of the parameters are derived from the likelihood and interpreted as describing the frequency with which different parameter values would be observed if the same experiment were repeated a large number of times.
In the Bayesian approach, the likelihood is combined with \textit{prior} probability distributions for the model parameters (summarising what is known about the parameters before seeing the data) to derive the \textit{posterior} probability distributions for the parameters. These are interpreted as the degree of credence that should be given to the parameter values based on the current data and any previous knowledge.

The advantage of the frequentist approach is that it is objective (i.e., all researchers will agree on the $p$-value and confidence limits, whose interpretation is calibrated). The big disadvantage is that the ``objectivity'' is obtained by neglecting any prior knowledge about the properties being measured, which is waste of information.

The main advantage of Bayesian inference is that all inferences logically follow from Bayes' theorem (i.e., given a prior all calculations have the certainty of deductive logic). In fact, it provides a natural and principled way to incorporates relevant prior probabilities (i.e., when new observations become available, the previous posterior distribution can be used as a
prior). The main critique is that the results may strongly depend on the choice of the priors which are subjective. Since there is no single method for choosing a prior, different researchers can choose different priors and may therefore arrive at different posteriors and conclusions. Of course, one can try different priors (which is an advantage) and see how sensitive are results to the choice of priors, but that may come with a high computational cost, especially in models with a large number of parameters.

A growing number of Bayesian methodologies have been developed in
recent years to account for the various complications involved in
modelling cluster scaling relations (see, e.g.,
\citealt{2007ApJ...665.1489K}, \citealt{2010MNRAS.404.1922A},
\citealt{2016MNRAS.457.1279M}, \citealt{2016MNRAS.455.2149S}).

In most regression analyses, the intrinsic scatter is modelled as a normal distribution in log space (and this appears to be a good description of the data). The scatter is usually measured in one direction (i.e., vertically or horizontally) relative to the regression line (with the scatter in one direction related to that in the other by the slope of the line).
When natural logarithms are used, the standard deviation of the measured scatter can then be interpreted as the fractional or percentage scatter in one property relative to the other.
The quality of available data usually requires that the scatter is treated as a constant across the fitted range, but it is physically plausible that this is not the case; for example, low mass clusters should be more impacted by non-gravitational processes and may hence show more scatter in their properties.

\begin{figure}[t]
  \centering
  \includegraphics[trim={15cm 4cm 15cm 0},width=0.7\textwidth]{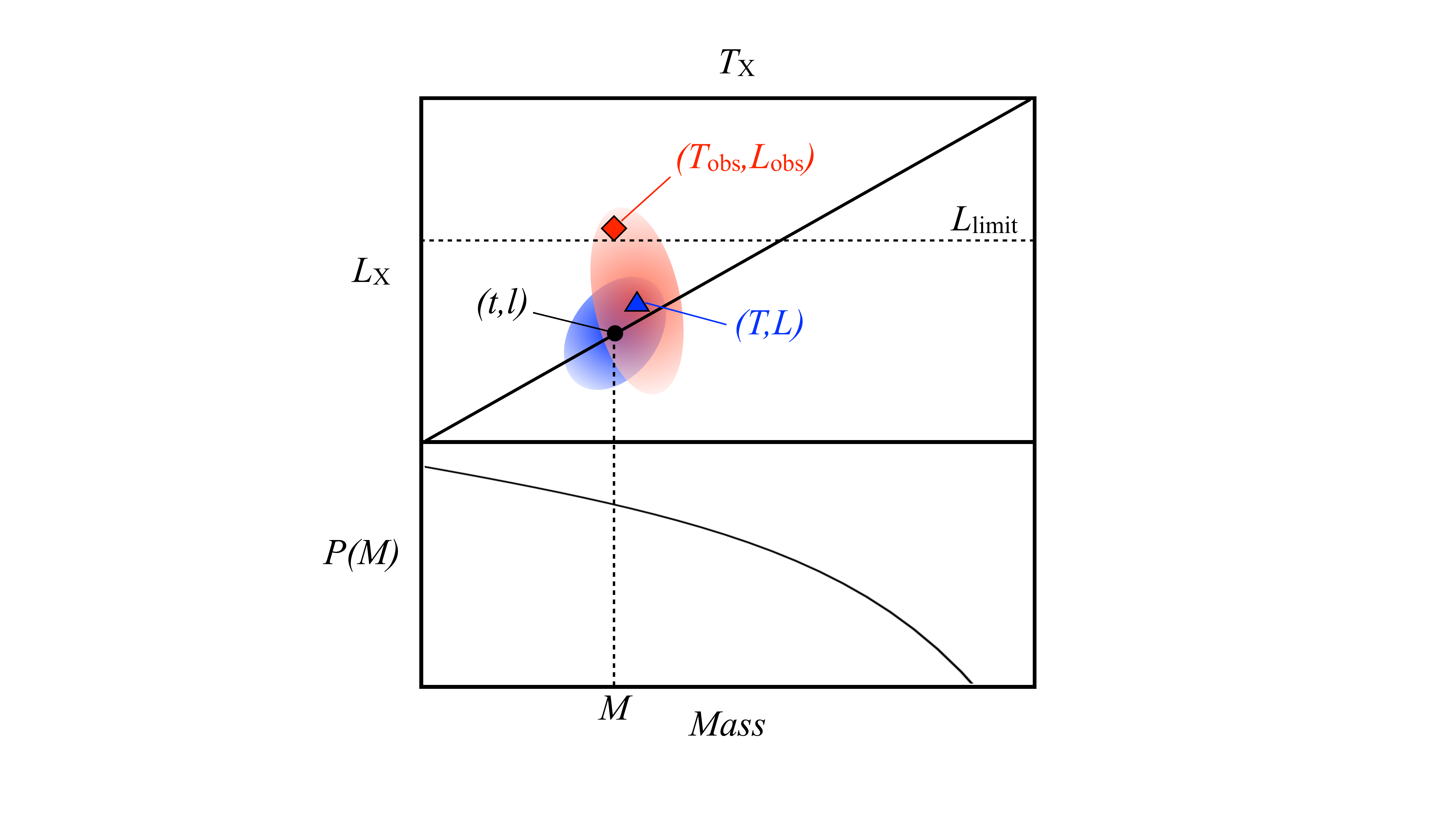}
  \caption{Visualisation of the likelihood of observing cluster
    properties given a model of scaling relations \citep[adapted
    from][]{2019A&A...632A..54S}. The lower panel illustrates the
    prior probability on cluster mass, based on a theoretical mass
    function. The scaling relations model predicts that a hypothetical
    cluster of mass $M$ would have luminosity $l$ and temperature $t$,
    placing it on the \LT\ relation at the black circle in the upper
    panel. The intrinsic scatter in $\Tx$ and $\Lx$ is illustrated by
    the blue ellipse, the tilt of which indicates covariance in the
    scatter. The blue triangle shows a possible location of the
    hypothetical cluster given this scatter. Uncertainties in the
    measurement of the observed quantities $T_\text{obs}$ and
    $L_\text{obs}$ are illustrated by the red ellipse, with the red
    diamond illustrating possible values of the observed cluster
    properties given the combination of intrinsic scatter and
    measurement error. Finally, the horizontal dashed line illustrates
    that a detection limit could be applied. In this case the
    cluster's observed luminosity is higher than $L_\text{limit}$,
    allowing it to be detected in spite of its relatively low mass.}
  \label{fig:bayreg}
\end{figure}

Figure \ref{fig:bayreg} illustrates the scenario we are dealing with when fitting the observed data. Let's consider the \LT\ relation, which is a manifestation of the correlations of $\Lx$ and $\Tx$ with mass. 
We can consider the true (i.e., unbiased) mass as the fundamental cluster property (denoted as $M$ in the figure). 
The true masses of a population of clusters will be described by the mass function, which serves as a prior on $M$ in the absence of any observational constraints. 
This is illustrated in the bottom panel of Figure \ref{fig:bayreg}.

For a hypothetical cluster of mass $M$, our scaling relation model
predict the ``theoretical'' temperature $t$ and luminosity $l$ of the cluster given a set of scaling relation parameters (slope,
normalisation, evolution). 
The ``true'' values $T$ and $L$ that would be measured given an observation of infinite accuracy and precision are related stochastically to $t$ and $l$ due to the intrinsic scatter (illustrated by the blue ellipse in Figure \ref{fig:bayreg}). 
The ``observed'' cluster properties $T_\text{obs}$ and $L_\text{obs}$ are, in turn, related stochastically to $T$ and $L$ by the measurement uncertainties (illustrated by the red ellipse in Figure \ref{fig:bayreg}). 
Finally, Figure \ref{fig:bayreg} illustrates a simple survey selection based on a threshold in $L_\text{obs}$. 
In this example, an object whose true properties lie below the survey threshold is included due to the combination of intrinsic and observational uncertainties.

The same approach can be applied to scaling relations between any combination of observables. 
The core of the Bayesian analysis methods is then the construction of a likelihood model that describes the stochastic relations between the observed, true and theoretical values in the presence of any selection criteria.

\subsection{Multivariate analysis}\label{sect:multifit}
The study of relationships between cluster property pairs provides a valuable but fundamentally limited view of the data. 
In fact, there is far more information that can be extracted by jointly fitting the form of the scaling relations between multiple observables. 
For example, the \LT\ relation is often studied, but in the self-similar model, this two-dimensional correlation is actually a manifestation of the correlations of \Mgas\ and \Tx\ with mass.

A (sometimes overlooked) consequence of combining scaling relations in the derivations of the self-similar relations in Section \ref{sec:self-similarity} is that their parameters become linked.
For example, since we combined the \TM\ and \MgM\ relations to derive the \LM\ relation, the slope and evolution of the \LM\ relation depends on the slope and evolution of those more fundamental relations.
Departures from self-similar behaviour in either or both of the \TM\ and \MgM\ relations would thus lead to non-self-similar scaling of luminosity with mass.

Similarly, the slope and evolution of the \LT\ relation depends on those of the more fundamental scaling relations. 
In particular (and perhaps counter-intuitively), the algebra leads to a dependence of the expected \textit{evolution} of the \LT\ relation on the \textit{slope} of the \LM\ and hence \TM\ and \MgM\ relations. 
These dependencies can be straightforwardly taken into account when jointly modelling multiple cluster properties in a multivariate analysis (\citealt{mau14}).

Multivariate analyses are becoming more common and can provide important information about the physics of the ICM, removing ambiguity that may exist in the interpretation of a simple two-dimensional scaling relation (e.g., does a cluster fall above the \LT\ relation because it is brighter or cooler, or both?).
For instance \cite{mau14} showed that the observed ICM scaling relations can be described by a deficit of gas in low-mass clusters compensated for by higher gas temperatures such that the total thermal energy of the gas remain constant for a given mass.
This could be the result of AGN feedback removing low entropy gas while heating the remaining gas, an effect working more efficiently at low masses.
\cite{2015MNRAS.446.2629E} showed that the deviations from self-similarity can be explained by three physical mass-dependent
quantities: the gas clumpiness, the gas fraction, and the slope of the thermal pressure. 
\cite{2020MNRAS.492.4528S} found evidence that the change in slopes is due to radiative cooling and AGN feedback.

In a multivariate analysis, the intrinsic scatter of the cluster properties is generally modelled as a multivariate normal distribution, parameterised by a covariance matrix. 
The off-diagonal terms of the covariance matrix then describe the degree of correlation between the scatter of the individual properties. 
This encodes useful astrophysical information on the processes responsible for the scatter. 
As noted in \textsection \ref{sec:observational-biases}, the covariance must also be modelled to correctly capture the effects of selection biases.

Several studies have started to measure the covariance of X-ray properties (and those at other wavelengths), finding non-zero covariance between various quantities (e.g., \citealt{2014MNRAS.441.3562E}, \citealt{2014MNRAS.438...78R}, \citealt{mau14}, \citealt{2015MNRAS.446.2205M}, \citealt{2019A&A...632A..54S}). \cite{2012MNRAS.426.2046A} argued that these correlations are due to internal structure, cluster orientation, formation history, and uncorrelated structures.

A multivariate cluster dataset can also be considered as a cloud of points which can assume any form of structure, not just linear
correlations parallel to the axes. 
Therefore, one can use principal component analysis (PCA) to find a small number of scaling relations (i.e., pair of cluster properties) to describe most of the variance in the dataset with a small number of new uncorrelated parameters. 
PCA finds the direction in the parameter space where the data are most elongated using least-squares to minimize the variance (the first principal component). 
The second component is then found by searching for the most elongated direction after the first component is removed, and so forth. 
Thus, the first principal component provides the smallest distance to a line in the space of the original variables, the second provides the smallest distance to a line in the plane perpendicular to the first component, and so on. 
Therefore, when the cluster properties lie on a plane defined by the first two principal components, the scaling relations on this plane are defined by the third principal component (\citealt{1984MNRAS.206..453E}).

In a related approach, \cite{2012MNRAS.420.2058E} found that all the standard self-similar scaling relations occur as special cases of a generalized scaling relation with the form $M\propto A^{\alpha}B^{\beta}$. 
Here $A$ and $B$ respectively represent ICM properties relating to the depth of the halo gravitational potential (e.g., the gas temperature) and on the distribution of the ICM (e.g., the gas mass or X-ray luminosity) which is more affected by the physical processes determining the ICM global properties (e.g., AGN feedback). 
These generalised scaling relations were found to have significantly reduced intrinsic scatter compared with traditional two-dimensional scaling relations (\citealt{2012MNRAS.420.2058E},
\citealt{2013MNRAS.435.1265E}, \citealt{2018MNRAS.474.4089T}).

\subsection{X-ray telescope calibration}
Beside the selection biases and the differences in the fitting techniques, there is another issue complicating the comparison between different studies, namely the calibration uncertainty between different instruments.

The calibration of X-ray instruments is challenging, because there are no absolute calibration targets (i.e., sources whose absolute flux and spectral properties are known) and because of the uncertainties in the overall calibration, which relies on a combination of ground-based calibration measurements and in-flight calibration programs. Ground-based calibrations are not perfect because of the time and budget pressure during the mission development phase, and because of technical limitations in the calibration facilities. In-flight calibrations are complicated as well because the instruments performance can change with time due to many different reasons (e.g., radiation damage,  filter deterioration, component failures, contamination
buildup, etc).

In the last decade the International Astronomical Consortium for High Energy Calibration\footnote{\url{http://web.mit.edu/iachec/}} supported several studies on clusters of galaxies in order to investigate the effective area cross-calibration status between different X-ray instruments.
\cite{2010A&A...523A..22N} found that in the hard band (2-7 keV) the energy dependence of the effective area of various X-ray telescopes were accurately calibrated but the normalisation of the effective areas disagreed by 5–10\%, with the disagreement becoming significantly larger in the soft energy band (0.5–2.0 keV).

These calibration differences lead to a systematic difference in the ICM temperature determined with different instruments. 
For instance, \citet[see also \citealt{2010A&A...523A..22N}]{2015A&A...575A..30S} found that the cluster temperatures obtained with \XMM\ are systematically lower than those obtained with \Chandra. 
The situation is further complicated by the temperature dependence of this difference. 
In the low temperature regime the differences are relatively small (a result that seems to hold also when including {\it  Suzaku} data; e.g., \citealt{2013A&A...552A..47K}), but for high temperature clusters the difference can reach 20\% or more.

This has an obvious impact on the scaling relations studies. 
For instance, it is obvious that an analysis based on temperatures measured with \Chandra\ would produce a flatter \LT\ relation than would be obtained by using \XMM\ derived temperatures. 
Other relations are impacted as well. 
In fact, due to the temperature-dependence of the discrepancy, this issue has also an impact on the gradient of the temperature profiles used to calculate the total mass via the HE equation. Therefore, one need to be careful when comparing the scaling relations determined with different instruments.

\subsection{Emission-weighted and spectroscopic-like temperatures}
A further complication in the measurement of the temperature of the ICM arises because it is not isothermal.
Due to the limitation of the current instruments, a single temperature model is often used to fit the observed X-ray spectra.
However, the observed spectrum is the result of an integral of emission along the line-of-sight through the ICM, which will include multiple components with different temperatures (and metallicities).
A single temperature fit will then yield an ``emission weighted'' temperature (i.e., an average temperature that is weighted by the relative brightness of the different thermal components). The telescopes and CCD detectors used in current observatories have a larger effective are in the soft band.
This means that a single-temperature fit will be further weighted towards the lower temperature components of the ICM that contribute the most emission in softer bands (e.g., \citealt{2004MNRAS.354...10M}, \citealt{2006ApJ...640..710V}).

A related problem arises when comparing the thermal structure of observed and simulated clusters. To derive a single temperature for a simulated cluster, one must assign some weighting to each temperature component in its ICM. The resulting ICM temperature can be written as
\begin{equation}\label{eq:T}
    T=\frac{\sum_i w_i T_i}{\sum_i w_i}
\end{equation}
where $T_i$ is the temperature of the $i$ component of the ICM, which contributes with a weight $w_i$.
By assuming that $w_i$=$m_i$ (where $m_i$ is the mass of ICM in the $i$ component) one obtains $T_\text{mw}$, the so-called mass–weighted temperature.
Instead, by weighting the different temperatures with the radiative emission contributions of the different phases (i.e.,  $w_i$=$\epsilon_i$), one obtains $T_\text{ew}$, the emission weighted temperature (e.g., \citealt{2001ApJ...546..100M}). However, \cite{2004MNRAS.354...10M} showed that the fitted temperature from {\it Chandra} or {\it XMM-Newton} spectra is biased low with respect to $T_\text{ew}$ and introduced $T_\text{sl}$, the spectroscopic–like temperature which is recovered by Equation \ref{eq:T} using the weight
\begin{equation}\label{eq:sl}
    w_i=\rho_i m_i T_i^{-0.75} .
\end{equation}
The spectroscopic–like temperature was generalized to the case of  arbitrary metallicities by \cite{2006ApJ...640..710V}. 
The two methods provide very similar results in the high temperature regime (i.e., in the limit where spectra are continuum-dominated) but significant variations may exist in the low-temperature regime because of the effects on the temperature introduced by the metallicity dependent line emission. 
The form of the weights given in Eq. \ref{eq:sl} was obtained for CCDs having a band-pass and a spectral resolution like those on board {\it Chandra} and {\it XMM-Newton}. 
Such resolution blurs the emission lines and do not allow them to be clearly separated from the continuum. 
It is likely that the weighting scheme in Eq. \ref{eq:sl} will not be appropriate for high-resolution spectrometers planned for future missions, and revised weightings will be required.

\section{Observational results and deviations from self-similarity}
\label{sec:obs-scaling}

One of the main reasons to study the scaling relations of galaxy clusters is to test for departures from the self-similar expectations.
These would then give insight into the astrophysical processes that are not included in the self-similar models, and be used to test the predictions of more realistic cluster models such as those from numerical simulations. Understanding the behaviour of the scaling relations is also essential for their use in cosmological studies. The key predictions of the self-similar model that are tested observationally are that the slopes of the observed scaling relations follow the self-similar predictions (sometimes called \textit{strong} self-similarity), and that the normalisation follows the predicted evolution (sometimes called \textit{weak}
self-similarity). 
The scatter and covariance in scaling relation may also be measured, giving astrophysical insight while also being crucial for the application of scaling relations to estimate cluster
masses from observable properties.

In the following, we will review the observational constraints on the slope, evolution, and scatter of the various X-ray scaling relations.
We will also consider results on the use of ICM scaling relations for mass estimation.
In doing so we will make a distinction between studies that compare X-ray properties of the ICM with ``direct'' and
``indirect'' mass measurements.
Direct measurements are those which measure the mass of a cluster without needing calibration, such as hydrostatic or weak lensing masses.
Indirect measurements are those which rely on calibration of some observable quantity against direct mass measurements.

When considering the scaling relation between an X-ray observable and indirect masses, one is implicitly including a second scaling relation between mass and some other observed quantity that was used to derive the indirect masses.
Thus, we prefer scaling relations measured using direct masses, but these are harder to construct (and hence more scarce) as direct mass estimates are observationally expensive to obtain.
Unless stated otherwise, the scaling relations we discuss are
based on direct mass measurements, and properties are measured within $\Rf$.

\subsection{The slopes of scaling relations}
\label{sec:slop-scal-relat}

\begin{figure}[t]
  \centering
  \includegraphics[width = \textwidth]{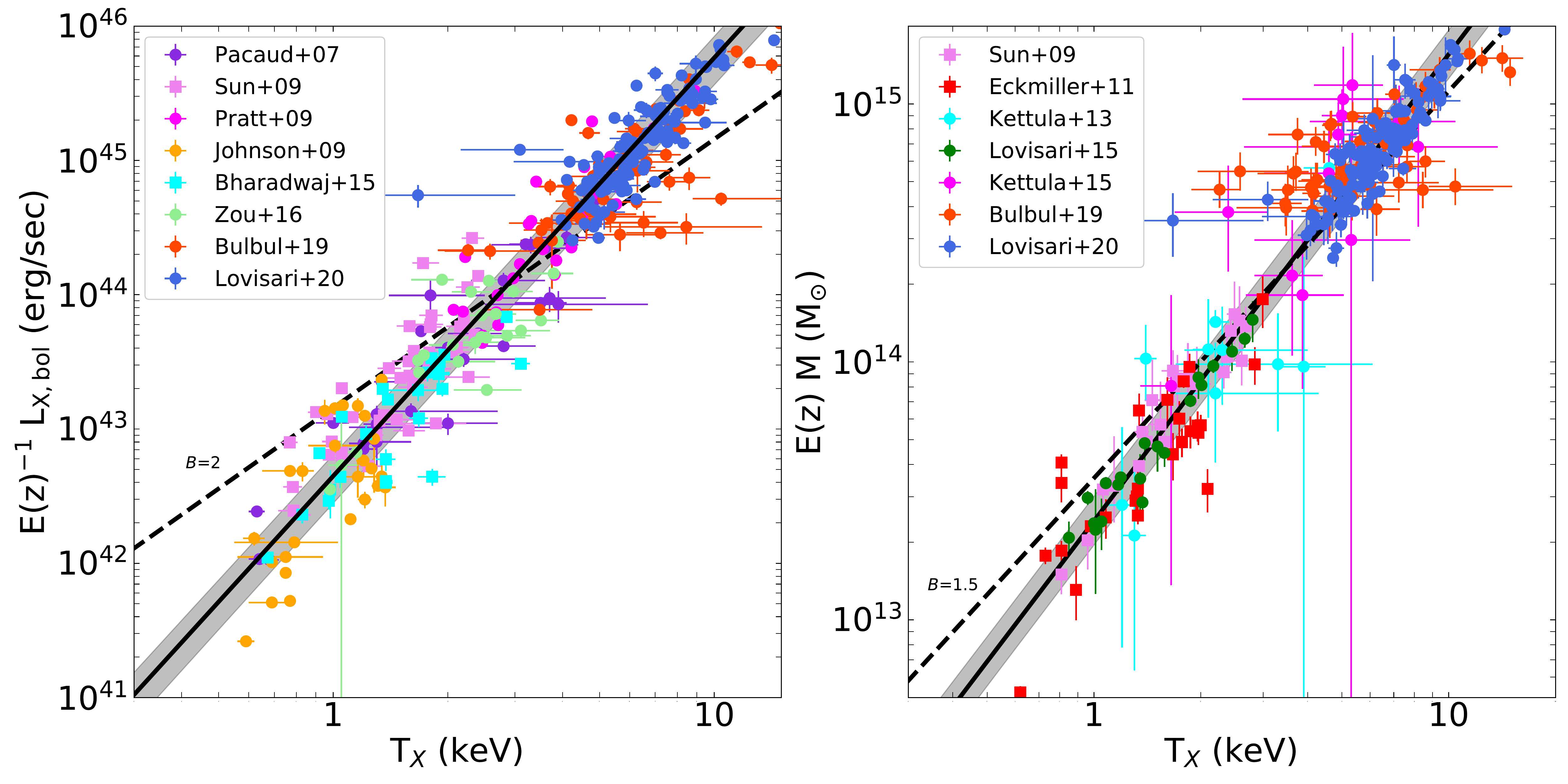}
  \caption{Figures illustrating compilation of \LbolT\ and  \MT\ measurements; this is adapted from \citet{lov21}. The {\it Chandra} measurements (squares) are converted to  {\it XMM-Newton}-like temperatures using the relations given in \citet{2015A&A...575A..30S}. The solid lines represent the fitted relations (with the 1$\sigma$ uncertainties indicated by the shaded areas) including all systems, and are compared with the case predicted by the self-similar scenario (dashed). The fits have been performed with the LIRA Bayesian regression package (\citealt{2016MNRAS.455.2149S}).}
  \label{fig:lt-compilation}
\end{figure}

The most widely studied of the ICM scaling relations is that between $\Lx$ and $\Tx$, due to the fact that both quantities are relatively easy to determine from shallow X-ray observations. 
The earliest observations of this scaling relation suggested that the slope was steeper than the self-similar expectation (\citealt{mit79}), and these results have been borne out in a large number of subsequent studies.
While the exact value of the slope found varies (likely due to e.g. differences in modelling, sample definition and selection effects, instrumentation and calibration), a broad consensus finds $2.5 \lesssim \Blt \lesssim 3.5$ for the bolometric \LT\ relation (\citealt{gio13}). Figure \ref{fig:lt-compilation} shows a compilation of measurements of the \LT\ relation from the literature. 
This illustrates that the observed \LT\ relation is well-described by a power law over a wide range of temperatures and luminosities, but the slope of the relation is incompatible with the slope predicted by the self-similar model.

There is some evidence that, when their core regions are excluded, the most massive and relaxed galaxy clusters do exhibit a self-similar \LT\ relation (\citealt{mau12}). 
The outer parts of these relatively simple systems seem to conform to the assumptions of the self-similar model and be largely free from the influence of non-gravitational effects. 
This is not unreasonable, since the central regions of clusters are where non-gravitational processes such as radiative cooling and energy input from AGN and stellar processes are strongest (see \textsection \ref{sec:interpretation}).

The determination of the \LT\ relation in the lower mass regime of galaxy groups ($\Tx\lesssim 2 \keV$) is challenging, with a wider range of observational results, with some evidence that the relation steepened further for galaxy groups (e.g., \citealt{osm04}). 
This is likely due to the larger scatter in ICM properties for groups combined with the difficulty in forming large samples with good quality data due to their intrinsic faintness. 
More recently, the quality of samples has improved, allowing for analyses that model selection biases (\textsection \ref{sec:observational-biases}), and a picture has emerged in which lower-mass systems follow the same \LT\ relation as high mass clusters, albeit with larger intrinsic scatter (\citealt{lov21}). However, given the increasing contribution from line emission to the X-ray luminosity for cooler systems (\textsection \ref{sec:lm-relation}), we would expect the \LT\ relation to flatten for galaxy groups. 
This may imply that a steepening of the \LT\ relation for galaxy groups is being masked by the contribution of line emission.

The deviation of the \LT\ relation from the self-similar predictions can also be explained in terms of entropy variation. Since the overall level of entropy in the ICM is set by the mass of the cluster (through the strength of the accretion shock), then a given entropy increase will have a stronger effect on the ICM in lower-mass clusters. 
This would have the effect of modifying the slopes of the scaling relations involving the mass, density and luminosity of the ICM compared with the self-similar model. 
This is confirmed observationally with the cores of galaxy groups exhibiting a higher entropy than is achievable by pure gravitational collapse (see \citealt{2021Univ....7..142E} and references therein) which then leads to a flatter $K$--\Tx\ relation (e.g., \citealt{Pratt2010}). 
However, the slope of the $K$--\Tx\  relation is found to agree better with the self similar value when the inner regions are removed from the analysis (e.g., \citealt{sun09}) indicating a connection with the feedback processes at play in the core of galaxy systems.

Observational studies of the scaling of ICM properties with mass are more challenging due to the difficulty in estimating cluster masses. 
In the case of the \LM\ relation, the overall picture from the literature (based on both direct and indirect mass measurements) is that the slope is significantly steeper than the self-similar expectation, with $1.5\lesssim \Blm \lesssim 2$ typically found for bolometric luminosities (where $\Blm=4/3$ is expected). 
This agrees qualitatively with observations of the \LT\ relation; the luminosities of low mass (i.e., low temperature clusters) are lower than expected in the self-similar model.

Given that the self-similar \LT\ and \LM\ relations are derived from the \MgM\ and \TM\ relations, the fact that the luminosity scaling relations depart from self similarity imply that one or both of $\Tx$ and $\Mgas$ must also depart from self-similar scaling with mass.
Observations of the \TM\ relation broadly find slopes that roughly agree with the self-similar expectation of $\Btm=3/2$ (see right panel of Figure \ref{fig:lt-compilation}), although care must be
taken when comparing temperatures with hydrostatic masses since those are derived from the same temperature measurements.
On the contrary, observations of the \MgM\ relation consistently favour a slope steeper than $\Bmgm=1$, which can be interpreted straightforwardly as a lower gas fraction in less massive clusters.
However, when analyses are restricted to high-mass systems ($\Mf\gtrsim 5 \times 10^{14}\Msol$), the gas fraction appears to become constant (i.e., $\Bmgm=1$), implying self-similar behaviour of the most massive clusters (\citealt{man16}). 
A constant \fgas\ for more massive systems implies that the scaling relations may ultimately be best described by a broken power law, with self-similar slopes where \fgas\ is constant, and a steeper slope at lower masses. However, current observations tend to be well described by a single power law.

A similar argument can be used when investigating the \YM\ relation.
In most studies the slope of this relation is found to be slightly steeper (although often consistent within the uncertainties) than what is predicted from the self-similar scenario. 
However, given the steeper slope of \MgM\ relation, this is not surprising.

Collectively, observations show that the luminosity of the ICM has a steeper scaling with mass or temperature than predicted by the self-similar model, due to the fact that lower mass systems contain less X-ray emitting gas than predicted.
The temperature of the gas, meanwhile, obeys a scaling close to that expected for an ICM that is in virial equilibrium with the cluster potential.
The principle physical process at work must therefore be to remove the ICM from the potential well of lower mass systems (or equivalently inhibit its accumulation into those systems; \textsection \ref{sec:non-grav-proc}).

\subsection{The evolution of scaling relations}
\label{sec:evol-scal-relat}

The evolution of cluster scaling has proved difficult to measure due to the challenge of obtaining good quality data for sufficient numbers of distant clusters with a well-understood selection function, along with defining a consistent low-redshift baseline against which to measure the evolution. 
Inconsistencies in sample definition, analysis methods or instrumentation between local and distant samples could create (or obscure) evolution in the data.
Given these limitations, attention has been focussed on testing the self-similar prediction of an evolving normalisation of the scaling
relations. 
In principle the slope and scatter may also evolve (not predicted by the self-similar model), but measuring these is not currently feasible given difficulty of observing the ICM in distant clusters with current observatories.

The \LT\ relation is again the best-studied of the scaling relations in this regard, but no consensus exists on its evolution. Measurements in the literature encompass all possibilities from negative evolution ($\glt<0$) to evolution that is stronger than self-similar.
\citet{rei11} performed a meta-analysis of samples of clusters out to $z\approx 1.5$, finding evidence for negative evolution but highlighted the influence that the choice of local sample can have on the inferred evolution.

The most reliable evolution measurements come from analyses of large samples spanning a wide redshift range, and which model the effects of selection biases on the scaling relations. 
These are now becoming available (e.g., \citealt{gil16}, \citealt{bul19}, \citealt{lov20}). 
Thus far, these do not provide any strong evidence that the evolution of the \LT\ relation differs from the self-similar prediction, but the statistical uncertainties remain significant.

The evolution of the mass-observable scaling relations is even less well-constrained than the \LT\ relation, due to the difficulty in performing mass measurements for distant clusters. 
There is some variation in the measurements reported, with the most precise constraints tending to come from small heterogeneous samples for which the systematic uncertainties are largest. 
A few studies have been performed with large, self-consistent samples including modelling of selection biases, and in general these show no strong evidence for evolution that differs from self-similar expectations for any of the mass-observable scaling relations (\citealt{man16}, \citealt{lov20}, \citealt{chi21}), although \cite{ser21} have recently found some hints for weaker than self-similar evolution of the \TM\ relation.

The evolution of the scaling relations (particularly involving \Lx) is complicated by the fact that the cores and outer parts of cluster may plausibly be expected to evolve differently. 
It is clear that non-gravitational processes, to which the prediction of self-similar evolution does not apply, are most important in the centres of clusters. 
Indeed, the prevalence of the strongly centrally peaked ICM surface brightness profiles that we associate with cool cores is found to be lower at high redshift. 
This appears to be due to the cool cores being stable over cosmic time, while the ICM at larger radii evolves self-similarly, reducing in brightness towards the present time which in turn makes the cores more pronounced (\citealt{mcd17}).

One might expect to see differences in the evolution of \Lx\ scaling relations when the core regions are, or are not, excluded from the luminosity measurement. 
In fact, this is difficult to test, since it is not possible to self-consistently exclude the core regions and model the selection biases for an X-ray selected cluster sample, since the core regions significantly contribute to the detection probability of a cluster. Studies using SZ-selected clusters have found no indication that evolution in the \Lx\ scaling relations differed when cores were excluded, but the uncertainties remain large (\citealt{bul19},  \citealt{lov20}; with the former based on indirect mass measurements). 
There is some limited evidence that the scatter in the \Lx\ scaling relations is reduced for distant clusters (\citealt{mau07b}, \citealt{man16}). 
This is a plausible consequence of the relative decrease in the strength of cool cores compared to the surrounding ICM in more distant clusters. 
The scatter may also be reduced by non-cool-core clusters being missed in samples of high redshift systems due to their low contrast against the background. 

Overall, current observations do not indicate that the evolution of galaxy cluster scaling relations is different from self-similar expectations out to $z\lesssim 1.5$. This implies that clusters have effectively formed (or regained equilibrium after their most recent disturbance) recently compared to the redshift at which they are observed. There is no strong evidence for astrophysical evolution of clusters beyond that due to the use of overdensities in the self-similar model. However, the uncertainties remain large, and evolution that is weaker or stronger than the self-similar predictions cannot be confidently excluded.

\subsection{Scatter and covariance}
\label{sec:scatter-covariance}

The non-zero intrinsic scatter in scaling relations is one very clear way in which they depart from self-similar predictions.
Observational constraints on the scatter (and covariance) in ICM scaling relations can be used to probe the physical processes responsible.

The scatter in scaling relations involving \Lx\ is found to be significant, with e.g. typically $\sim50\%$ scatter found in \Lx\ at fixed \Tx. 
However, this is dominated by the variation in core properties of clusters, with cool-core clusters having significantly enhanced luminosities (\textsection \ref{sec:interpretation}).

\begin{figure}[t]
  \centering
  \includegraphics[width = 0.49\textwidth]{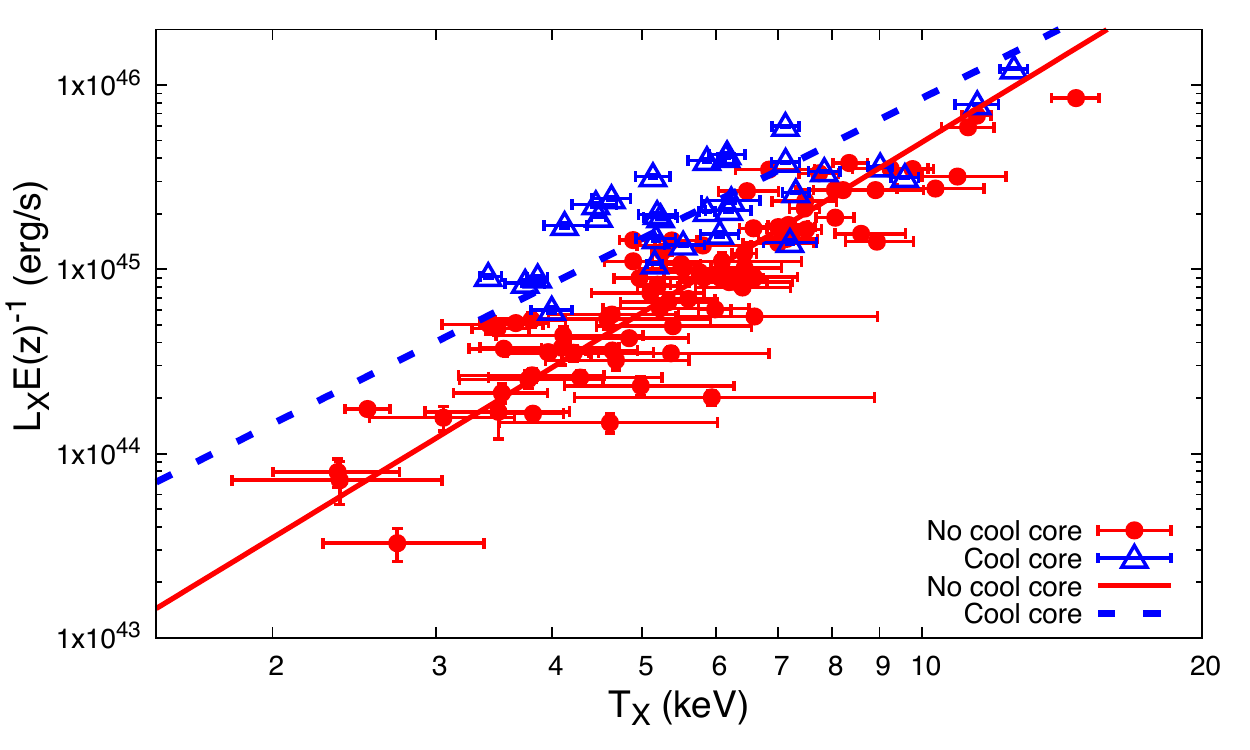}
  \includegraphics[width = 0.49\textwidth]{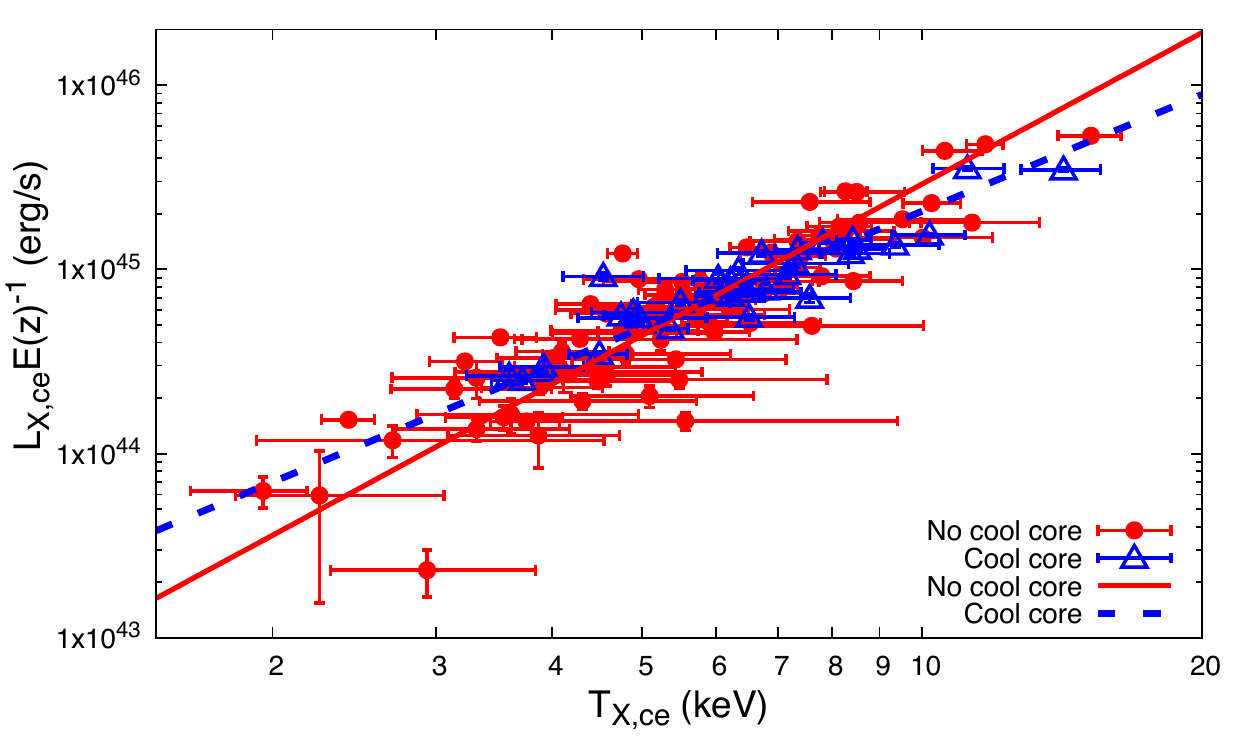}
  \caption{Figure illustrating impact of cool cores on the scatter in
    the \LT\ relation, adapted from \citet{mau12}. In both plots, blue
    points indicate clusters classified as possessing cool cores,
    while red points denote those without cool cores (the lines show
    the best fits to the subsamples). In the left panel, luminosity
    and temperature are measured within an aperture of radius $\Rf$.
    In the right panel, the central $15\%$ of $\Rf$ was excluded from
    the luminosity and temperature measurements. For this sample,
    excluding the core regions reduces the scatter in luminosity from
    about $70\%$ to about $30\%$.}
  \label{fig:lt-cc}
\end{figure}

Exclusion of the core regions (conventionally the central $0.15\Rf$) from the measurements significantly reduces the scatter to of order 10--20$\%$ (with the largest reduction for the most relaxed clusters).
This is illustrated in Figure \ref{fig:lt-cc}, in which clusters are separated into those hosting a strong cool core and those that do not.
The cool core clusters are clearly offset, but once the core regions are excluded, all clusters occupy a similar part of the \LT\ plane.

The most insight into the physical origin of the scatter in the scaling relations is gained by measuring the covariance of observables with respect to mass. This can be assessed by a multivariate analysis in which the mass-observable relations are modelled simultaneously (see \textsection \ref{sect:multifit}). Such studies are becoming more common (e.g., \citealt{mau14}, \citealt{far19a}), and one of the most robust results is a strong positive covariance between \Mgas\ and \Lx. 
This is unsurprising given the dependence of \Lx\ on \Mgas\ (Equation \eqref{eq:l-rho}): clusters with a higher than average mass of ICM relative to their total mass will be correspondingly more luminous in X-rays.

\begin{figure}[t]
  \centering
  \includegraphics[width = 0.5\textwidth]{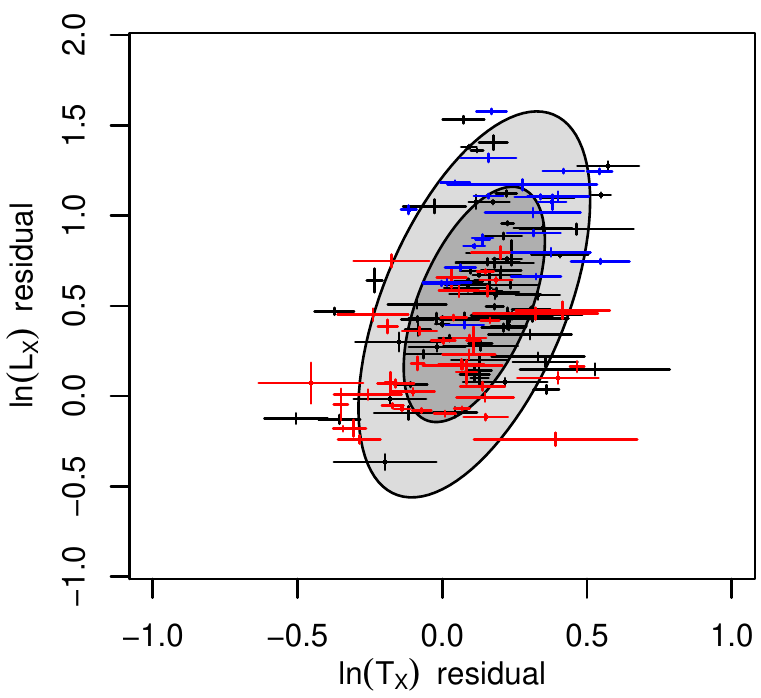}
  \caption{The correlation between the residuals in \Lx\ and \Tx\
    relative to their mass scaling relations. The blue and red
    datapoints show the most relaxed and disturbed clusters
    respectively. The shaded ellipses illustrate the best fitting
    covariance, with the dark and light regions indicating the $1\sigma$ and $2\sigma$ intervals, respectively, of the measured scatter. Figure adapted from \citet{man16}.}
  \label{fig:lt-resid}
\end{figure}

Most observations have also found a positive covariance between \Lx\ and \Tx\ with respect to mass (i.e., clusters which have higher than average luminosities for their mass tend also to have higher than average temperatures, and vice-versa). 
This is illustrated in Figure \ref{fig:lt-resid}, which shows the correlation between the residuals in \Lx\ and \Tx\ relative to their mass scaling relations. 
There is a positive correlation, which is associated with the dynamical state of the clusters. 
Numerical simulations show that the positive covariances between X-ray properties probably stands at any redshift (e.g., \citealt{2010ApJ...715.1508S}, \citealt{2018MNRAS.474.4089T}).

The most relaxed clusters (blue points in Figure \ref{fig:lt-resid}) tend to be both hotter and more luminous than average, which is driven by the association of cool cores with relaxed clusters, which significantly enhance their luminosity while having a smaller impact on their overall temperature. Clusters undergoing a merger (red points), meanwhile, tend to be cooler and less luminous than average. 
The impact of mergers on ICM properties is complex, with shocks temporarily and locally boosting both luminosity and temperature. However, once a merger starts, the mass of the cluster can be considered to have increased instantly, but the temperature and luminosity of the ICM will not reach the final values expected for that mass until sufficient time has passed for the gas to become virialised and settle in the new potential (a few Gyr). 
Until this point, the temperature and luminosity will be lower than average for the new mass.
Minor mergers can also contribute to the scatter because smaller and colder substructures are efficiently stripped and mixed due to stellar and AGN feedback (\citealt{2018MNRAS.474.4089T}).

The covariance of X-ray properties with those measured at other
wavelengths can give further insight. There have been attempts to
measure the covariance between \Mgas\ and the total stellar mass,
\Mstar, in member galaxies of a cluster. The measurements are
challenging, since cluster masses are required in order to provide the
scaling relations about which the covariance is measured, and
observational results are mixed. There is some evidence for negative
covariance between \Mgas\ and \Mstar\ (\citealt{far19a}). This can be
understood if clusters are considered to be gravitationally closed,
with a fixed proportion of their mass in baryons. A cluster of a given
mass that had formed more of its baryons into stars would have a
correspondingly smaller mass of baryons in the ICM.

\subsection{Mass proxies}
\label{sec:mass-proxies}
A key application of cluster scaling relations is to infer the masses
of clusters. For this to be done accurately, the slope and
normalisation of a mass-observable scaling relation must be
calibrated, which is generally accomplished by measuring direct
masses for a subset of the clusters. Ideally this would be done for a representative sample of clusters (i.e. subject to the same selection biases as the parent sample) and in the
mass and redshift range in which the mass proxies were to be applied. However, often this is not possible, and e.g. a self-similar form for
the mass-observable relation is assumed when applying the mass proxies
at redshifts beyond those for which the proxies were calibrated.

Even if the form and evolution of a mass-observable scaling relation
were known perfectly, the intrinsic scatter about the relation would
limit the precision with which masses could be estimated. Significant
effort has thus been devoted to identifying observables (or their
combinations) which possess the minimum intrinsic scatter with
mass\footnote{While it is conventional to report the scatter in a
  particular mass proxy at a given mass, the primary quantity of
  interest for mass estimation is the scatter in mass at a given value
  of the proxy. The two scatters are directly related by the slope of
  the scaling relation.}. The calibration of a
low-scatter mass proxy against direct masses for as large and
representative a sample of clusters as possible would then enable mass
estimates for large numbers of clusters.

Simply measuring the scatter between an X-ray property and a direct
(or indirect) mass measurement can be misleading, since this will include
contributions from the scatter of both the X-ray property and the
mass measurement with the true mass.

Hydrostatic masses give the most precise direct mass measurements, but
they are only secure for relaxed clusters for which the assumption of
hydrostatic equilibrium is valid. However, there is not a unique
approach to derive the hydrostatic masses (see, e.g.,
\citealt{2013SSRv..177..119E} and \citealt{2019SSRv..215...25P} for
advantages and disadvantages of the different methods), and the
results are not always in good agreement.

Weak lensing masses are agnostic to the dynamical state of the cluster, but have more limited precision with an estimated $\sim30\%$ scatter between the weak lensing mass and true mass, due primarily to projected structures and non-sphericity of the systems (\citealt{bec11}). 
Calibrating a low-scatter proxy via its scaling relation with weak lensing mass would improve the accuracy of the proxy while avoiding the large scatter in weak lensing masses, and this is the most promising way to proceed.

As seen in Sect \ref{xoptsz} the SZ effect probes the hot gas in the ICM independently from the X-ray observations. Since the integrated SZ flux, $Y_{\rm SZ}$\footnote{This is related to the Compton parameter $y=(\sigma_T/m_e c^2)\int P dl$, which is a measure of the thermal electron pressure of the ICM gas along the line of sight, where P is the ICM thermal electron pressure, $\sigma_T$ is the Thomson cross section, $m_e$ is the electron rest mass, and c is the speed of light. $Y_{\rm SZ}$ is then given by the integral of $y$ over the solid angle of the cluster.},  from a cluster directly measures the total thermal energy of the gas (and thereby of the total mass),  it is expected to provide a low-scatter mass proxy. This expectation is confirmed by both numerical simulations (e.g., \citealt{2004MNRAS.348.1401D}) and indirectly from X-ray observations using \Yx~ (i.e., the X-ray counterpart of the SZ signal (e.g., \citealt{2021ApJ...914...58A}).

X-ray luminosity is the most easily measured property of the ICM, and can be recovered from low signal-to-noise data if the cluster redshift is known. 
However, unless core regions are excised, \Lx\ has the largest scatter with mass of the X-ray properties of the ICM. 
It is also strongly influenced by non-gravitational physics, which is apparent in the non-self-similar slope of the \Lx\ scaling relations. 
This also means it is plausible that the evolution of the \LM\ relation may differ from self-similar expectations (although this is not yet strongly tested observationally). 
All of this makes \Lx\ a sup-optimal mass proxy. 
However, the use of \Lx\ may be necessary if the resolution or data quality prohibit the exclusion of core regions (typically the case in survey data). 
Furthermore, X-ray selection of clusters is based on the entire emission, so the relation between \Lx\ and mass is needed for analyses that rely on modelling the selection function of such a sample in detail.

If the cores are excised, the luminosity (\Lce) becomes a much better mass proxy, with e.g. \citet{man16a,man18} estimating a scatter of $10-20\%$ relative to the true mass, based on comparisons with either hydrostatic or weak lensing masses, which include the scatter between those quantities and true mass (although \citealt{far19a} report a significantly larger scatter in \Lce\ relative to true mass, based on weak lensing masses). Relatively low scatters are also found when comparing \Lce\ with indirect mass measurements (e.g., \citealt{mau07b}, \citealt{pra09a}).

Historically, \Tx\ has been regarded as a useful mass proxy, due to
its simple connection to cluster mass via the virial theorem.
Comparisons of temperatures with hydrostatic masses for relaxed
clusters indicate that the scatter is low ($\lesssim10\%$
scatter in \Tx; \citealt{vik06a}, \citealt{arn07}, \citealt{sun09}, \citealt{man16a}). However, these analyses may
underestimate the scatter since they tend to neglect the covariance
between \Tx\ and hydrostatic mass due to the physical dependence of
the latter on the former, and also due to correlated measurement
errors if both quantities are determine from the same X-ray data.
Studies that utilise weak lensing masses instead, tend to measure
scatters that are a little larger (typically $10-20\%$ in
\Tx; \citealt{man16}, \citealt{far19a}, \citealt{chi21}).

The mass of the ICM, \Mgas, has been found to be have a scatter with
mass that is similar to, or smaller than that of \Tx. \Mgas\ has the
additional advantage as a mass proxy in that it can be measured
reliably with lower quality data than \Tx. For relaxed clusters with
hydrostatic masses, the scatter in \Mgas\ relative to mass is found to
be $\lesssim 10\%$ (\citealt{arn07}, \citealt{man16a}). Studies based on weak lensing
masses again tend to report slightly higher scatters of $10-20\%$
(even when modelling the scatter between lensing mass and true
mass; \citealt{far19a}, \citealt{chi21}).

The final X-ray mass proxy we will consider is \Yx, which is expected
to be a low-scatter tracer of cluster mass due to its relation to the
total thermal energy content of the ICM. This is because \Mgas\ and
\Tx\ tend to react in opposite ways to non-gravitational processes,
with the effects approximately cancelling to reduce the scatter of
$\Yx$ with mass. In case of mergers the gas mass increases while the
measured temperature may be lower due to the presence of colder
structures or incomplete virialisation of the ICM. In case of AGN
feedback, the gas mass may decreases because it is expelled by the AGN
activity which at the same time is heating the gas. Observational
studies based on hydrostatic masses support this, finding a scatter
that is comparable to (or a little smaller than) \Mgas\ (i.e., $\lesssim10\%$; \citealt{arn07}, \citealt{sun09}). Once again, measurements utilising
weak lensing masses tend to find higher scatter, spanning a range from
about $10\%$ to $30\%$ (e.g., \citealt{mah13}, \citealt{far19a}, \citealt{chi21}). This could
be due in part to covariance between hydrostatic mass and $\Yx$ that
is neglected in X-ray studies suppressing the scatter to some degree.

In summary, all of the X-ray properties of clusters have some use as
mass proxies, with strengths and weaknesses. If the data quality are
sufficient, then \Yx\ or \Mgas\ alone are probably the most precise
mass proxies. For lower quality data, \Lce\ is a good alternative. It
should also be noted that for low temperature clusters, the
computation of \Mgas\ becomes quite sensitive to the metal abundance
of the ICM, which cannot generally be well constrained in lower
quality data. Luminosities less strongly affected by this additional
systematic uncertainty (\citealt{eck20}).

Of course, uncertainties on all of the proxies increase at lower
masses and higher redshifts where they are largely untested. This is
principally due to the observational difficulties in obtaining direct
mass measurements in those parts of the parameter space.

\section{Interpretation of scaling relations}
\label{sec:interpretation}

It should come as no surprise that real clusters do not perfectly conform to the simplistic self-similar scaling relations. However, the self-similar model provides a useful reference with which to compare observed cluster scaling relations. As we have seen in \textsection \ref{sec:obs-scaling}, it provides a good description of some, but by no means all, of the observed scaling behaviour of clusters.

For instance, all the relations involving quantities strongly
connected with the gas distribution (i.e., luminosity, gas mass and therefore \Yx) show a steepening with respect to the self-similar expectations. 
On the contrary, the \MT\ relation is found to be largely consistent with the predictions. 
While there is a general consensus on the steeping of the relations, it is not yet clear if the evolution is weaker or stronger than the self-similar predictions.
Part of the reason is the poor sampling of groups and clusters at high redshift which lead to very large uncertainties in the estimation of $\gamma$, but the large intrinsic scatter can also easily hide weak effects of the evolution. 
Now, the obvious question is what are the possible mechanisms that could explain the observed relations in term of deviation from the self-similar slopes, scatter, and (lack of) evolution?

Departures from self-similarity are signs of astrophysical processes beyond those included in the simplifying assumptions underpinning the self-similar model. 
These are often referred to as ``non-gravitational'' processes because the self-similar properties of the ICM are dictated by the gravitational potential of the cluster, and can be broadly separated into the categories of cooling and heating of the ICM (see \textsection \ref{sec:non-grav-proc}).

There are a number of indications that these non-gravitational
processes have changed the properties of the gas, particularly in poor clusters and groups (e.g., \citealt{2021Univ....7..142E}). 
For instance, due to the increasing complexity of the emissivity function in the low-temperature regime, in absence of any feedback the \LT~ and \LM~ relations are expected to gradually flatten at low temperatures and masses, contrary to observations. 
When comparing relations obtained in the group and cluster regimes, one must account for the increasing contribution of the line emission at low-temperatures, otherwise the impact of feedback processes could be underestimated.

Since the observed cluster X-ray properties do not agree very well with the self-similar predictions (i.e., scaling relations for purely gravitational heating), extra physics is required for a proper description of the ICM. 
As discussed in \textsection
\ref{sec:non-grav-proc}, a heating source is therefore required to raise the entropy of the ICM. This heating could happen before or after the cluster formation. 
If the non-gravitational heating occurred just prior to the cluster collapse, then the amount of heat needed is a few keV per particle (e.g., \citealt{2000ApJ...532...17L}). 
This minimum entropy in the pre-collapse intergalactic medium is expected to break self-similarity at the scale of groups.

An alternative to heating is that cooling could effectively remove low entropy gas from the centres of groups and clusters. This low entropy gas has the shortest cooling time, and once it has cooled sufficiently it will no longer emit in the X-ray band. This means that the average temperature (and entropy) of the gas that remains visible in X-rays will be increased (e.g., \citealt{2005RvMP...77..207V}).

\cite{1999Natur.397..135P} was the first to show that the entropy does not scale linearly with temperature (as expected from self-similar scenario). 
Later studies, confirmed that poor systems shows a significant excess to that achievable by pure gravitational collapse (e.g., \citealt{sun09}, \citealt{2014MNRAS.438.2341P}). 
Also the observed entropy profiles appears to be inconsistent with purely gravitational heating, showing a significant flattening in the cores (e.g, \citealt{2003MNRAS.343..331P}, \citealt{sun09}).

Currently, the favored mechanism is represented by energy release by central AGNs which are thought to heat and displace the thermal gas changing the radial and global properties of the clusters. 
It is an inside-out process which affects more significantly the regions at small radii. 
Moreover, the effect is expected to increase at lower masses impacting the shape of the scaling relations. 
The injection is likely provided by outflows from AGN (i.e, jets of plasma inflate bubbles as large as $\sim 100$~kpc in radius). 
This is sometimes referred to as the ``kinetic'' mode of AGN feedback (\citealt{fab12}). 
In this mode, outflows of relativistic material interact with the surrounding ICM through inflating bubbles and by driving shocks through the ICM. 
The shocks heat the ICM directly, while bubbles initiate motions in the ICM that they displace as they rise buoyantly,
which are then dissipated into heating via turbulence. 
The observed energetics of such bubbles are sufficient to maintain cool cores in an approximately steady state (\citealt{mcn07}). 

This kinetic-mode feedback is likely to shape the scatter of the scaling relations (particularly those involving luminosity), by virtue of its influence on cool cores. However, the feedback appears sufficiently well regulated that there is not excess energy that would remove gas from low mass systems and hence modify the slopes of scaling relations.

Instead, the slopes of scaling relations are likely to be set by radiative-mode AGN feedback operating at high redshift (peaking at around $z\sim 2-3$). 
This is a form of pre-heating, increasing the gas entropy at early stages in the growth of clusters. 
Many AGN will be located in galaxies that will ultimately reside at the centres of groups and clusters of galaxies as the surrounding structures collapse. 
The energetic impact of AGN in this mode is primarily in radiation coupling to the surrounding gas and dust, and either heating or pushing the gas sufficiently that it escapes from the dark matter halo in which the AGN resides (such that it is not re-accreted again over the lifetime of the cluster). 
This feedback acts on the lowest entropy gas closest to the AGN, effectively removing it from the halo, and leaving behind higher-entropy gas that takes longer to cool and settle to the centre of the halo (\citealt{mcc11}). 
This effect is naturally stronger in lower mass halos whose gravitational potentials are less able to hold on to the heated gas (or equivalently, a smaller entropy increase is needed to prevent the gas sinking to the inner regions). 
The hot gas content of these systems is thus reduced, altering the slopes of the ICM scaling relations (e.g., \citealt{2014ApJ...783L..10G}).

\subsection{Comparison with simulations}\label{hydsim}

The comparison between the observed scaling relations with the results from hydrodynamical simulations offer a great opportunity to constrain the physical processes at work in the ICM and to enable the investigation of the selection effects in surveys of groups and clusters. 
In simulations one has the advantage of being able to turn on and off different physical processes (e.g., galactic winds, AGN feedback) keeping the same initial conditions. 
By over-plotting the results obtained from different simulations with the observed scaling relations it is possible to obtain hints about the processes which are driving the evolution of groups and clusters. 
In fact, it is frequent to obtain a significant offset between observed and simulated relations (e.g., \citealt{2018MNRAS.474.4089T}). This could reflect a different gas distribution (over- or under-peaked), or a different masking in observations and simulations of substructures present within \Rf\ (since there is not a unique definition of what constitutes a distinct substructure).

For instance, \citet{2008ApJ...687L..53P} showed that simulations
including only stellar feedback were unable to reproduce the observed
\fgas--\Tx\ relation. However, by adding AGN feedback they were able
to reduce the gas fraction within \Rf, and match the observational
values. The intensity of the AGN feedback regulates the amount of gas
that is ejected from the potential well. The role of the AGN feedback
in reproducing the observed properties of galaxy clusters have been
confirmed in many other studies (e.g., \citealt{2010MNRAS.406..822M},
\citealt{2010MNRAS.401.1670F}, \citealt{2014MNRAS.441.1270L},
\citealt{2014ApJ...783L..10G}, \citealt{2017MNRAS.465.2936M}). It is
possible to obtain reasonable gas mass fractions with models without
AGN feedback, but the \fgas--\Tx\ relation is flatter than observed
(see the reviews by \citealt{2021Univ....7..142E} and
\citealt{2021Univ....7..209O} for more details). Reproducing the
correct \fgas--\Tx\ (or the \MgM) relation is crucial for a proper
comparison between observations and simulations. The gas fraction is
found to be mass-dependent (e.g.,
\citealt{vik06a}, \citealt{2007ApJ...666..147G},
\citealt{pra09a}, \citealt{2015A&A...573A.118L}, \citealt{2016A&A...592A..12E}; see also the
reviews by \citealt{2021Univ....7..142E} and
\citealt{2021Univ....7..209O}). A lower gas fraction leads to lower
luminosity and \Yx\ parameter impacting the shape of the relations.
Therefore, as highlighted in \cite{lov20} the mass range of the
investigated sample plays a significant role in both the slope and
normalization of the derived scaling relations.

The gas fraction of low mass systems is an extremely sensitive to the feedback scheme implemented in simulations (see \citealt{2021Univ....7..209O}). However, in general when feedback processes are included in the simulations the derived scaling relations are in better agreement with the observed properties. This confirms the gravitational processes are insufficient to fully describe the ICM properties.

\section{Summary and Future outlook}
Until the 90s the largest cluster samples with X-ray data were mostly drawn from optical surveys. In most cases the samples were completely inhomogeneous and the selection function unknown.
A big step was achieved with {\it ROSAT} that detected a few thousand galaxy clusters in its all sky-survey, and allowed the construction of flux-selected samples of several tens to hundreds of clusters (e.g.,  HIFLUGCS, \citealt{2002ApJ...567..716R}; REXCESS, \citealt{2007A&A...469..363B}).
More recently, the SZ surveys provided new catalogs of clusters which have been followed-up in X-ray.

In most cases the samples contained of order of one hundred objects, many of them in the local Universe, with only a few objects above redshift 1. In the near future, surveys across the electromagnetic spectrum will produce samples of tens of thousands objects out to redshift 2-3.

In the optical and near-infrared bands, {\it Nancy Grace Roman Space Telescope}, {\it Vera Rubin Observatory}, and {\it Euclid} will provide cluster detections and precise redshift and weak-lensing measurements to complement studies of the ICM properties.

Next-generation SZ surveys (e.g., the {\it Simons observatory}, {\it CMB-S4}) are forecast to detect over 1,000 clusters at $z>2$. This will allow studies into the epoch of cluster formation and an evolutionary timeline that are inaccessible with current catalogs.

Meanwhile, large radio surveys (e.g., {\it LOFAR}, {\it MWA}, {\it SKA}) can also provide useful information to investigate the dynamic and thermodynamic structures observed in the other wavelengths.
Moreover, the detection of radio halos and relics can provide information on the acceleration of particles during cluster mergers providing further insights into the virialization process.

Studies of the X-ray scaling relations will be advanced by dedicated X-ray surveys along with follow-up observations of the ICM in distant clusters detected at other wavelengths.

\subsection{eROSITA}
The {\it extended ROentgen Survey with an Imaging Telescope Array}
({\it eROSITA}, \citealt{2021A&A...647A...1P}) is operating at L2 orbit on-board the 'Spectrum-Roentgen-Gamma' satellite.
{\it eROSITA} consists of seven co-aligned X-ray mirror modules covering the soft band (i.e, 0.2-10 keV), with a large field of view and moderate spatial resolution.
Thanks to the combination of these characteristics {\it eROSITA} is 20–30 times more sensitive than the {\it ROSAT} sky survey in the soft band and is providing the first all sky imaging survey in the hard band (i.e., 2-10 keV).
{\it eROSITA} is planned to perform a 4-yr-long all-sky survey (started in December 2019) and is expected to detect $>$10$^5$ groups and clusters with mass \Mf$>$10$^{13}$M$_{\odot}$ out to redshift $\sim1$ (median: $z\sim$0.3; e.g., \citealt{2012MNRAS.422...44P}, \citealt{2014A&A...567A..65B}). 
This large number of clusters will allow to significantly improve  the constraints of several cosmological parameters.  
However, these goals will be achieved only if we will be able to improve significantly our knowledge of the \LM\ (and other) scaling relation.
Most of the systems detected with {\it eROSITA} will have too few photons to measure temperature profiles (\citealt{2014A&A...567A..65B}), and so the masses required for cosmology will need to be estimated through scaling relations.

\subsection{ATHENA}
To be launched in the 2030s, the {\it Advanced Telescope for High ENergy Astrophysics} ({\it ATHENA}) will comprise two instruments:
the X-ray Integral Field Unit (X-IFU; \citealt{2016SPIE.9905E..2FB}) and the Wide Field Imager (WFI; \citealt{2013arXiv1308.6785R}).
While the former will allow spectroscopic investigation with unprecedented resolving power the latter will offer unrivaled survey power (over small to medium sized fields) with X-ray imaging with  good spatial resolution. The WFI survey is expected to detect 10$^4$ objects at $z>$0.5 including a few tens of clusters at $z>2$, and measure their temperature to better than 25\%
accuracy (e.g., \citealt{2020A&A...642A..17Z}).

Perhaps the most exciting aspect of {\it ATHENA} will be its ability to study the ICM properties in follow-up observations of the most distant clusters detected in other surveys.
With such observations it will be possible to transform the currently poor constraints on the evolution of the scaling relations and to determine the history of the physical processes that dominate the injection of non-gravitational energy in the ICM.

\section{Acknowledgements}
The authors thank A. Mantz and S. Ettori for providing an adapted
version of previous published figures. We also thank S. Ettori for
useful comments and suggestions that helped to improve and clarify the
presentation of this work. LL acknowledges financial contribution from
the contracts ASI-INAF Athena 2019-27-HH.0, ``Attività di Studio per
la comunità scientifica di Astrofisica delle Alte Energie e Fisica
Astroparticellare'' (Accordo Attuativo ASI-INAF n. 2017-14-H.0), and
from INAF ``Call per interventi aggiuntivi a sostegno della ricerca di
main stream di INAF''. BJM acknowledges support from the Science and
Technology Facilities Council (grant number ST/V000454/1).


\bibliographystyle{aa}
\bibliography{./sl}

\end{document}